\documentclass[pdflatex,sn-mathphys-num]{sn-jnl}

\usepackage{mathrsfs}
\usepackage{graphicx}%
\usepackage{multirow}%
\usepackage{amsmath,amssymb,amsfonts}%
\usepackage{amsthm}%
\usepackage{mathrsfs}%
\usepackage[title]{appendix}%
\usepackage{xcolor}%
\usepackage{textcomp}%
\usepackage{manyfoot}%
\usepackage{booktabs}%
\usepackage{algorithm}%
\usepackage{algorithmicx}%
\usepackage{algpseudocode}%
\usepackage{listings}%
\usepackage[utf8]{inputenc}
\usepackage{textcomp}
\usepackage[nolist]{acronym}
\usepackage{xspace}
\usepackage[version=4]{mhchem}
\usepackage{newunicodechar}
\usepackage{lmodern}
\newunicodechar{⁺}{\textsuperscript{+}}
\newcommand{\erthr}{\ce{Er^{3+}}\xspace}
\newcommand{\Ilow}{\ce{${}^\text{4}\text{I}_{\text{15/2}}$}\xspace}
\newcommand{\Ihigh}{\ce{${}^\text{4}\text{I}_{\text{13/2}}$}\xspace}
\begin{acronym}
    \acro{RE}[RE]{rare earth}
	\acro{PL}[PL]{photoluminescence}
	\acro{PLE}[PLE]{photoluminescence excitation}
	\acro{ESR}[ESR]{electron spin resonance}
        \acro{ODMR}[ODMR]{optically detected magnetic resonance}
        \acro{RE}[RE]{rare-earth}
        \acro{MBE}[MBE]{molecular-beam epitaxy}
        \acro{EL}[EL]{electroluminescence}
        \acro{CR}[CR]{count rate}
        \acro{DCR}[DCR]{dark count rate}
	\acro{SSPD}[SSPD]{superconducting single-photon detector}
	\acro{SOI}[SOI]{silicon-on-insulator}
	\acro{QIP}[QIP]{quantum information processing}
	\acro{VLSI}[VLSI]{very large-scale integration}
	\acro{CMOS}[CMOS]{complementary metal-oxide-semiconductor}
	\acro{SNR}[SNR]{signal-to-noise ratio}
	\acro{AOM}[AOM]{acousto-optical modulator}
	\acro{EOM}[EOM]{electro-optical modulator}
	\acro{HEMT}[HEMT]{high-electron-mobility transistor}
        \acro{ZEFOZ}[ZEFOZ]{zero first-order Zeeman}
        \acro{CMOS}[CMOS]{complementary metal-oxide-semiconductor}
	\acro{FinFET}[FinFET]{fin field-effect transistor}
        \acro{PECVD}[PECVD]{plasma-enhanced chemical vapour deposition}
        \acro{PL}[PL]{photoluminescence}
	\acro{FWHM}[FWHM]{full width at half maximum}
	\acro{RF}[RF]{radiofrequency}
        \acro{TOF}[TOF]{tunable optical filter}
        \acro{CW}[CW]{continuous-wave}
\end{acronym}

\theoremstyle{thmstyleone}%
\theoremstyle{thmstyletwo}%

\theoremstyle{thmstylethree}%

\begin{document}

\title[Article Title]{Narrow magneto-optical transitions in Erbium implanted silicon carbide-on-insulator}


\author[1]{\fnm{Alexey} \sur{Lyasota}}\equalcont{These authors contributed equally to this work.}
\email{a.lyasota@unsw.edu.au}
\author[2,3]{\fnm{Joshua} \sur{Bader}}\equalcont{These authors contributed equally to this work.}

\author[4]{\fnm{Shao Qi} \sur{Lim}}

\author[5]{Brett C. Johnson}

\author[4]{\fnm{Jeffrey C.} \sur{McCallum}}

\author[6]{Qing Li}

\author*[1]{\fnm{Sven} \sur{Rogge}}
\email{s.rogge@unsw.edu.au}
\author*[2]{\fnm{Stefania} \sur{Castelletto}}
\email{stefania.castelletto@rmit.edu.au}

\affil[1]{Centre of Excellence for Quantum Computation and Communication Technology,
School of Physics, University of New South Wales, Sydney, NSW 2052, Australia}

\affil[2]{School of Engineering, RMIT University, Melbourne, 3000, VIC, Australia}

\affil[3]{Centre for Quantum Computation and
Communication Technology, School of Engineering, RMIT University, Melbourne, 3000, VIC, Australia}
\affil[4]{Centre for Quantum Computation and Communication Technology, School of Physics,
The University of Melbourne, Melbourne, 3010, VIC, Australia}
\affil[5]{School of Science, RMIT University, Melbourne, 3001, VIC, Australia.}
\affil[6]{Electrical and Computer Engineering, Carnegie Mellon University, Pittsburgh, 15213,
PA, USA}

\abstract{

Solid-state spin–photon interfaces operating in the near-telecom and telecom bands are a key resource for long-distance quantum communication and scalable quantum networks. However, their optical transitions often suffer from spectral diffusion that hampers the generation of coherent spin–photon entanglement.
Here we demonstrate narrow magneto-optical transitions of erbium dopants implanted into thin-film silicon carbide (SiC)-on-insulator, a viable platform for industrially scalable quantum networks. Using high-resolution resonant spectroscopy and spectral hole burning at cryogenic temperatures, we reveal sub-megahertz homogeneous linewidths and identify two lattice sites that best stabilise the emitters. We further characterise their optical lifetimes and magneto-optical response, establishing erbium-doped SiC-on-insulator as a robust and scalable platform for on-chip quantum networks.

}




\maketitle
\section*{Main}\label{sec1}
Solid-state spin–photon interfaces \cite{esparza2011high,gao2015coherent,awschalom2018quantum,kindem2020control} operating in the telecom band offer a direct route to long-distance quantum communication and quantum networks \cite{kimble2008quantum, wehner2018quantum,ranvcic2018coherence,JonghoonExtended2024,Zhou_quantum_networks2025,gritsch2025optical}. In particular, spin–photon interfaces consisting of a point defect or impurity, interfaced to photons via spin-selective optical transitions \cite{wolfowicz2021quantum}, such as color centres in diamond \cite{pompili2021realization, parker2024diamond}, silicon \cite{redjem2020single}, silicon carbide \cite{Zhou_quantum_networks2025}, 2D-materials \cite{stern2022room} and rare earth ions \cite{ranvcic2018coherence,gritsch2025optical}, are candidates for the physical nodes of quantum networks.  Their performance, however, is limited by spectral diffusion, which broadens emission lines and hinders the generation of indistinguishable single photons \cite{morioka2020spin, ourari2023indistinguishable}, required for spin-photon entanglement \cite{knaut2024entanglement}. 

Silicon carbide (SiC) is a particularly attractive host for future quantum networks \cite{son2020developing,castelletto2022silicon,Zhou_quantum_networks2025}, as it combines metal-oxide-semiconductor (CMOS) compatibility, a wide bandgap (3.26 eV) and an optical transparency from the visible to the mid-infrared. In particular, spin-qubits associated with intrinsic defects in the hexagonal polytype 4H-SiC, such as the silicon vacancy \cite{widmann2015coherent} and divacancy \cite{christle2015isolated}, exhibit very long spin coherence times \cite{simin2017locking,nagy2019high,anderson2022five,zeledon2025minutelongquantumcoherenceenabled} and single photon indistiguishability \cite{morioka2020spin}. Over the last decade, 4H-SiC spin-photon qubits have accomplished key milestones, including electron spin control and entanglement with single nuclear spins \cite{bourassa2020entanglement}, electron spin single-shot readout via charge state control \cite{anderson2022five}, spin-photon entanglement \cite{FangSpin_Photon2024},  single photon emission within the telecom O-band \cite{cilibrizzi2023ultra} and coherent spin photo-electrical read-out \cite{nishikawa2025coherent}. Finally, 4H-SiC spin defects exhibit optical linewidths as narrow as 20 MHz \cite{anderson2019electrical}. These advances coupled with their compatibility with photonic integrated
circuits \cite{Crook2020,lukin20204h,PRXQuantum.1.020102,babin2022fabrication,day2023laser}, establish 4H-SiC as a leading solid state platform for
scalable quantum technologies.

Building on these achievements, thin-film 4H-SiC-on-insulator (4H-SiCOI) provides a scalable and industry-compatible route towards on-chip spin-photon interface quantum technologies based on low-loss integrated photonics \cite{lukin20204h,CaiOctave22,yang2023inverse,bader2024analysis,hu2024room,lipton2025low}.
Assessing the optical properties of candidate spin-photon emitters in this platform is therefore a crucial step towards SiC-based scalable quantum networks.

Among candidate emitters, the trivalent erbium ion (Er$^{3+}$) stands out for its optical transition near 1.5~$\mu$m. This is within the low-loss window of silica fibers and directly compatible with existing optical components developed at scale for telecommunication, minimizing losses in long-distance transmission with optical fibers for quantum networking \cite{ranvcic2018coherence}, and removing the need for frequency conversion \cite{dreau2018quantum}. While studies in bulk or heavily doped SiC have suggested clustering and spectral broadening effects, the potential of Er$^{3+}$ in 4H-SiCOI remains largely unexplored \cite{bader2025photoluminescence}.

Here, we demonstrate narrow magneto-optical transitions of low-fluence Er$^{3+}$ ions implanted in 4H-SiCOI. Using resonant photoluminescence excitation spectroscopy (PLE) and spectral hole burning at milli-Kelvin temperature, we resolve sub-MHz homogeneous linewidths and identify two distinct lattice sites that host stable emitters. 

\begin{figure}[H]
    \centering
    \includegraphics[width=\textwidth]{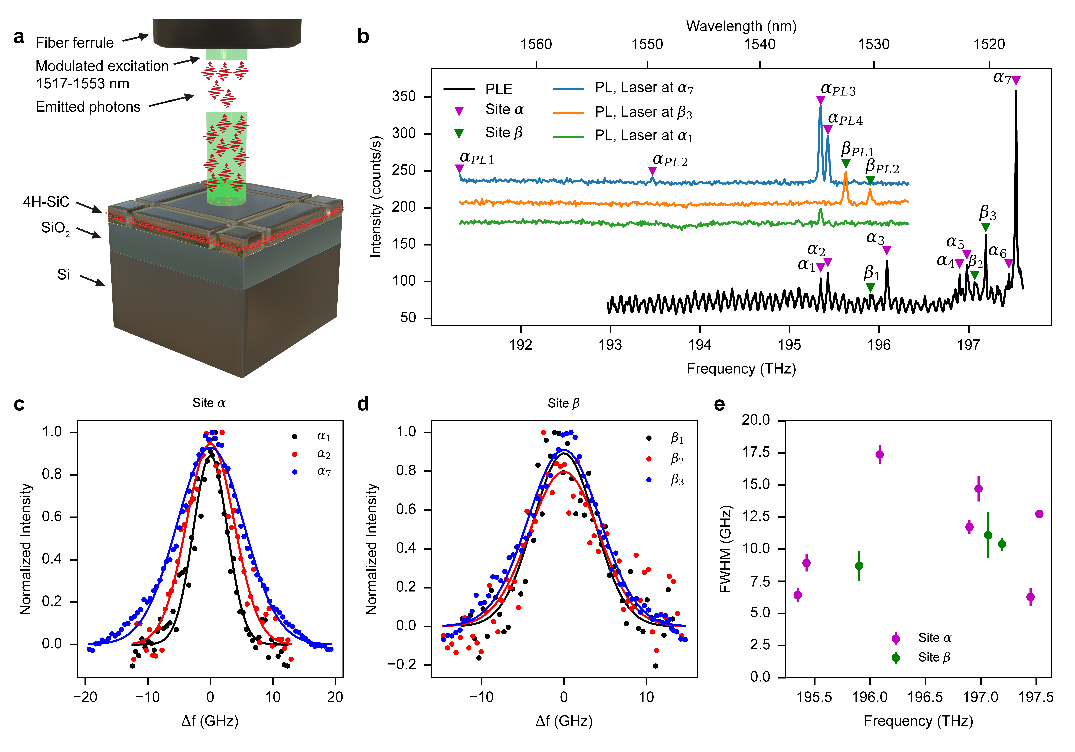}
    \caption{Er$^{3+}$-site identification utilizing \ac{PL}- and \ac{PLE}- measurements within 4H-SiCOI: \textbf{a} schematic of the experimental setup configuration where modulated tunable excitation (green) is applied onto a 4H-SiCOI-sample via a fiber ferrule (black), aiming to address the Er$^{3+}$-defects (red) embedded in the centre of a trenched 4H-SiC-layer; \textbf{b} \ac{PLE}- and \ac{PL}-spectra obtained from the Er$^{3+}$ implanted sample; 
    \textbf{c} identified narrowest optical transitions ($\alpha_{1}$, $\alpha_{2}$, $\alpha_{7}$) from Er$^{3+}$-site $\alpha$; \textbf{d} identified optical transitions from Er$^{3+}$-site $\beta$. The solid lines in \textbf{c} and \textbf{d} are Gaussian fits to the data; \textbf{e} Overview of observed inhomogeneous linewidths from identified $\alpha$ and $\beta$ resonances in relation to frequency where error-bars denote uncertainties from Gaussian fits.}
    \label{fig1}
\end{figure}

\erthr ions were introduced into the 4H-SiCOI substrates using low-fluence ion implantation, followed by thermal annealing to optically activate the \erthr and repair implantation damage. The optical properties were studied at 20 mK using photoluminescence (PL) and PLE spectroscopy. Fig. \ref{fig1}\textbf{a} schematically illustrates the experimental configuration. Tunable excitation pulses are coupled to the sample through a fiber ferrule, and the emitted photons are collected and directed to a superconducting single-photon detector inside a dilution $^{4}\text{He}/^{3}\text{He}$ cryostat (see Methods).

In total, 19 distinct PLE lines were identified between 194.968 and 197.528 THz (see Supplementary Information Figs. S1–S3). 
Ten of these resonances, shown in Fig. \ref{fig1}\textbf{b}, can be attributed to two distinct \erthr sites. Within the 4H-SiC crystal field, \Ilow and \Ihigh \erthr levels are expected to split into eight and seven Kramer doublets, respectively~\cite{babunts2000properties, choyke1997crystalfield}. At cryogenic temperatures, only the lowest Kramers doublet of the \Ihigh manifold is populated, such that up to seven transitions to the split \Ilow levels are expected in the PLE spectrum.
The two dominant \erthr sites are here denoted as $\alpha$ and $\beta$. Their \ac{PLE} (\ac{PL}) lines are labeled as $\alpha_i$ or $\beta_i$ ($\alpha_{PLi}$ or $\beta_{PLi}$) according to transitions between \Ilow $\rightarrow$ \Ihigh (\Ihigh $\rightarrow$ \Ilow), respectively. Transitions belonging to the same \erthr site were confirmed by correlating their \ac{PL}-spectra with specific PLE resonances, using a narrow-band tunable filter (see Methods and Supplementary Information). Both observed sites exhibited narrow inhomogeneous broadening, and for the more dominant site, sub-MHz homogeneous linewidths were resolved for transitions from the \Ilow ground state to the two lowest \Ihigh crystal field levels.
Magneto-optical spectroscopy further revealed that both $\alpha$ and $\beta$ sites share a single g-tensor orientation. At millikelvin temperatures, the PLE spectra are dominated by transitions from the ground-state doublet to the crystal field–split \Ihigh levels, allowing reconstruction of the energy level structure. 
Only like-to-like transitions were observed, consistent with optical transitions of high cyclicity~\cite{raha2020optical}.
For example, resonant excitation at the $\beta_3$ transition frequency (197.192 THz) yields a \ac{PL} spectrum with two peaks $\beta_{PL1}$ and $\beta_{PL2}$ (see Fig. \ref{fig1}\textbf{b}), with the $\beta_{PL2}$ \ac{PL} peak matching the $\beta_1$ \ac{PLE} peak frequency of 195.9018 THz. The lower frequency $\beta_{PL1}$ \ac{PL} peak does not have a \ac{PLE} counterpart. This indicates that the $\beta_{PL1}$ \ac{PL} peak corresponds to the transition from the lowest \Ihigh energy level to one of the higher \Ilow crystal field energy levels, which is not populated at the sample temperature. This allowed us to attribute these specific resonances to the same $\beta$ \erthr site. A similar procedure was applied for the identification of other observed $\alpha$ and $\beta$ transitions (see Supplementary Information, Figs. S4 and S5). A complete crystal field is observed for site $\alpha$, while site $\beta$ provides three distinct resonances.

Interestingly, when $\alpha_{3-7}$ are addressed in a \ac{PL}-measurement (see Fig. \ref{fig1}\textbf{b} for \ac{PL} spectra excited at the $\alpha_{7}$ frequency and Figs S4, S6, and S7 in the Supplementary Information for other \ac{PL} spectra), two distinct lines $\alpha_{PL3}$ and $\alpha_{PL4}$ are observed at the frequencies matching $\alpha_{1}$ and $\alpha_{2}$ \ac{PLE} transitions. The \ac{PL} spectrum excited via the $\alpha_{1}$ \ac{PLE} transition showed a single dominant \ac{PL} line at the $\alpha_{1}$ frequency (Fig. \ref{fig1}\textbf{b}). This confirms that the $\alpha_{1}$ and $\alpha_{2}$ \ac{PLE} lines correspond to transitions from the lowest energy level in \Ilow to the two lowest crystal field energy levels within \Ihigh (See Fig. S8 of the Supplementary Information).
Additional lower frequency $\alpha_{PL1}$ and $\alpha_{PL2}$ lines with much weaker emissions were observed in the \ac{PL} spectra. These lines correspond to transitions to higher \Ilow crystal field levels from one of the two lowest \Ihigh crystal field energy levels (See Fig. S8 in the Supplementary Information).
The large intensity ratio of \ac{PL} lines $\alpha_{PL3-4}$ and $\alpha_{PL1-2}$ can be explained by high branching ratios of $\alpha_{PL3}$ and $\alpha_{PL4}$ optical transitions, which implies a nearly unity branching ratio for the $\alpha_{PL3}$ transition.

The determined inhomogeneous broadening \ac{FWHM} values are 
(6.44 $\pm$ 0.55) GHz, 
(9.0 $\pm$ 0.78) GHz 
and (12.69 $\pm$ 0.47) GHz, 
for $\alpha_{1,2,7}$, as shown in Fig. \ref{fig1}\textbf{c}. The observed \ac{PLE} resonances $\beta_{1,2,3}$ show a similar inhomogeneous broadening as the $\alpha$ lines of (8.64 $\pm$ 1.16) GHz, (10.33 $\pm$ 1.53) GHz and (10.66 $\pm$ 0.75) GHz (see Fig. \ref{fig1}\textbf{d}). An overview of all observed resonances inhomogeneous linewidths is provided in Fig. \ref{fig1}\textbf{e}. 
Besides \ac{PLE} lines of $\alpha$ and $\beta$, we observed a number of weaker lines. These lines could not be assigned to a particular site due to their low intensity but showed comparable inhomogeneous broadening as sites $\alpha$ and $\beta$ (see Supplementary Information). 

The identification of only two main \erthr-sites in 4H-SiCOI with the low bound on the yield of $3\%$ for the site $\alpha$ with well-defined g-tensor (see Methods) is very promising. While the yield of the observed \erthr site formation in other CMOS-compatible platforms like Si \cite{berkman2023observing, berkman2025long} or \ac{SOI}~\cite{weiss2021erbium,gritsch2022narrow} is not quoted, the yield value of $\approx$ 1.6\% can be estimated from the recent results on the coupling of \erthr ions with photonic cavities~\cite{gritsch2025optical} (see Methods). In addition, there are several possible g-tensor orientations for \erthr in Si, which further reduces the yield of a site with well-defined spin energy splitting in the magnetic field below 0.8\% (see Methods).

The inhomogeneous broadening of the site $\alpha$ \ac{PLE} lines did not increase monotonically with the transition energy (see Fig.   \ref{fig1}\textbf{e}). This cannot be explained by the homogeneous linewidth broadening due to phonon-mediated relaxation within \Ihigh, commonly seen for higher energy optical transitions of optically active rare-earth defects in various other semiconductors~\cite{liu2006spectroscopic, berkman2025long}. The phonon-mediated relaxation of the \Ihigh energy level corresponding to the $\alpha_{2}$ optical transition is comparable with the optical lifetime of this level since the $\alpha_{PL4}$ \ac{PL} and $\alpha_2$ \ac{PLE} transitions appear at the same energy. The observed variations in inhomogeneous broadening can be explained by the variations in permanent electric or magnetic dipoles of different \Ihigh crystal field levels resulting in the \Ihigh crystal field level-dependent quantum confined Stark shift or magnetic dipole-magnetic dipole interactions between \erthr ions.

\begin{figure}[H]
    \centering
    \includegraphics[width=\textwidth]{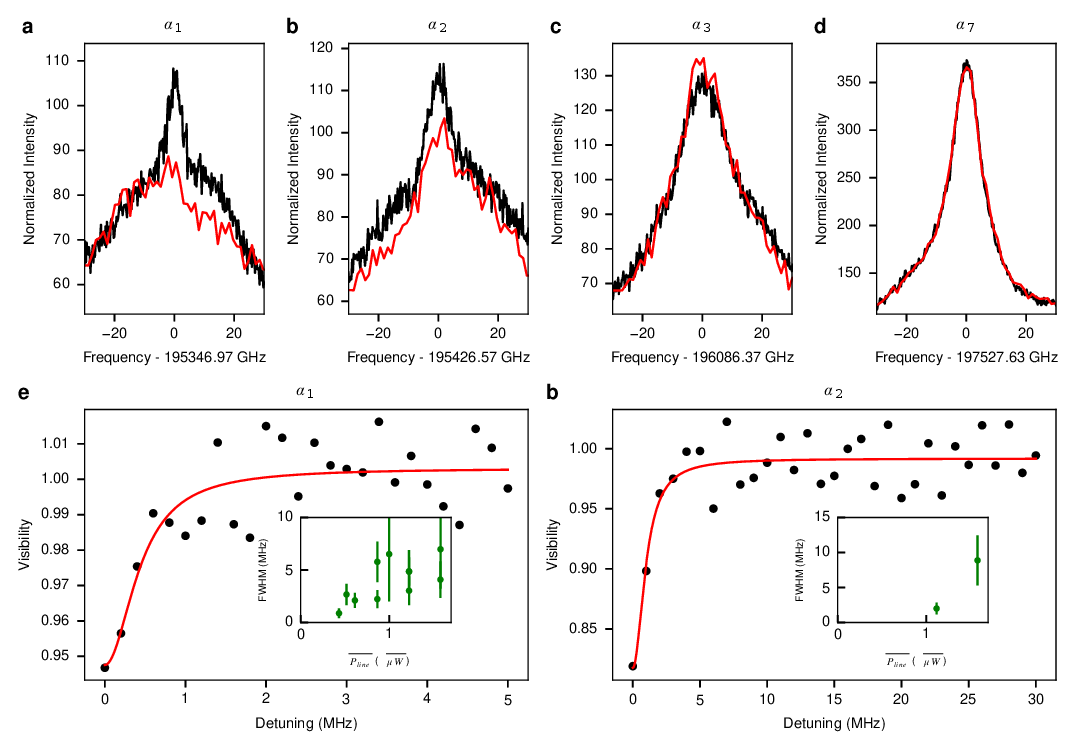}
    \caption{Spectral hole burning  and  homogeneous broadening: \textbf{a} Spectra of $\alpha_{1}$ optical transition measured with a frequency comb (black) and single laser configuration (red); \textbf{b} Spectra of $\alpha_{2}$ optical transition measured with a frequency comb (black) and single laser configuration (red); \textbf{c} Spectra of $\alpha_{3}$ optical transition measured with a frequency comb (black) and single laser configuration (red); \textbf{d} Spectra of $\alpha_{7}$ optical transition measured with a frequency comb (black) and single laser configuration (red); \textbf{e} Spectral hole burned in $\alpha_{1}$ fitted with a single Lorentzian fit (red) with $\text{P} = 0.19\ \mu\text{W}/\text{line}$. The inset illustrates the excitation power dependency $\sqrt{P_{line}}$ on the spectral hole (homogeneous) linewidth ranging between (0.88 $\pm$ 0.23) MHz to (6.97 $\pm$ 1.91) MHz ((0.44 $\pm$ 0.12) MHz to (3.49 $\pm$ 0.96) MHz)); \textbf{f} Spectral hole burned in $\alpha_{2}$ fitted with a single Lorentzian fit (red) with an inset illustrating the excitation power dependency $\sqrt{P_{line}}$ onto the spectral hole (homogeneous) linewidth ranging between (1.99 $\pm$ 0.42) MHz to (8.87 $\pm$ 1.81) MHz ((1 $\pm$ 0.21) MHz to (4.44$\pm$ 0.9) MHz)).}
    \label{fig2}
\end{figure}

We further examine multiple resonances from site $\alpha$ to provide insights into the coherence of the observable optical transitions, which is one of the key parameter for future quantum information processing applications. For that, we identified \ac{PLE} transitions with the longest optical coherence times (the narrowest homogeneous linewidth), determined by comparing \ac{PLE} spectra measured utilizing 60 MHz and sub-100 kHz optical excitations. The 60 MHz optical excitation was achieved by generating the optical frequency comb with densly spaced lines while sub-100 kHz optical excitation was achieved using a single laser frequency (see Methods). Absorption saturation \cite{berkman2025long} results in the large ratio between \ac{PLE} signal measured with sub-100 kHz and 60 MHz excitation bandwidth if the transition homogeneous linewidth is narrower than 60 MHz. Therefore, by comparing spectral measurements obtained with these two excitation parameters, we could determine transitions with narrow homogeneous linewidths. Fig. \ref{fig2}$\bf{a-d}$ show \ac{PLE} spectra of $\alpha_{1,2,3,7}$ lines obtained with these two excitation bandwidths. \ac{PLE} lines $\alpha_1$ and $\alpha_2$ have large signal ratios when measured using sub-100 kHz and 60 MHz excitation bandwidths identifying them as the prime candidates for the transitions with the narrow homogeneous linewidths (see Fig. \ref{fig2}$\bf{a}$ and $\bf{b}$). Higher energy transitions $\alpha_{3-7}$ did not show any significant difference in signal (see Fig. \ref{fig2} and Fig. S9) as expected due to the fast phonon relaxation rates between crystal field levels within the \Ihigh manifold \cite{berkman2025long}. 

Next, we performed transient spectral hole burning experiments \cite{berkman2023observing, berkman2025long, szabo1975observation, volker1989hole} to extract the homogeneous linewidths of the $\alpha$ and $\beta$ transitions, which provides further insights into their optical coherence (see Methods).
During this investigation, a frequency comb consisting of multiple optical doublets with detuning $\Delta f$ between two lines within the doublet is generated and applied to $\alpha_1$ and $\alpha_2$ inhomogeneous peaks. The frequency spacing between center frequencies of optical doublets $f_{comb}$ is kept constant and significantly exceeds the power-broadened \erthr optical linewidth.
By recording the emitted photons while tuning $\Delta f$, the spectral hole can be obtained where the spectral hole \ac{FWHM} is equal to twice the power-broadened homogeneous linewidth (see Fig. \ref{fig2}$\bf{e}$ and $\bf{f}$). The spectral hole and correspondingly homogeneous linewidths of $\alpha_1$ and $\alpha_2$ increase with the excitation power $P_{line}$, showing the expected $\sqrt{P_{line}}$ trend~\cite{berkman2025long}  derived from the detailed power-dependent measurements of the $\alpha_{1}$ spectral hole linewidth (see an inset in Fig. \ref{fig2}$\bf{e}$).

We identified a \ac{FWHM} of (441 $\pm$ 118) kHz and (1000 $\pm$ 215) kHz for resonances $\alpha_{1}$ and $\alpha_{2}$ at the lowest applied optical excitation power, as illustrated in Fig.\ref{fig2}\textbf{e} and \textbf{f}. These \ac{FWHM} values constitute the upper limits to the $\alpha_{1}$ and $\alpha_{2}$ optical homogeneous linewidth. The observed sub-MHz homogeneous linewidth deems the particular observable 
\erthr site resonance $\alpha_{1}$ among the narrowest transitions of any optically addressable emitter in SiC when being compared to silicon vacancies ($\text{V}_{\text{Si}}^{-}$) \cite{nagy2021narrow, heiler2024spectral}, divacancies ($\text{V}_{\text{Si}}\text{V}_{\text{C}}^{0}$) \cite{anderson2019electrical, he2024robust} and vanadium impurities (V$^{4+}$) 
\cite{cilibrizzi2023ultra, wolfowicz2020vanadium}.
The single spin S=1/2 V$^{4+}$, emitting in the telecom O-band from 1260 nm to 1360 nm, showed $\approx$ 100 MHz of spectral broadening in isotopically purified 4H-SiC, compared to several GHz in 4H-SiC with a natural abundance of isotopes~\cite{cilibrizzi2023ultra}, this being narrower of a single T-centre in \ac{SOI} (0.6-1 GHz)~\cite{higginbottom2022optical}. The single $\text{V}_{\text{Si}}^{-}$ (S=3/2) in natural abundance of nuclear spins 4H-SiC, emitting at 917 nm, showed a linewidth narrowing from 170 MHz down to 40 MHz only by forming a Schottky diode and applying an electrical control \cite{steidl2025single}. Single $\text{V}_{\text{Si}}\text{V}_{\text{C}}^{0}$ (S=1) in bulk intrinsic commercial material 4H-SiC, showed the narrowest linewidths between 130 to 200 MHz, and by applying electric fields, linewidths of 20 MHz were demonstrated~\cite{anderson2019electrical}; finally the modified divacancy PL6 has shown a linewidth of 720-820 MHz~\cite{he2024robust}.

\begin{figure}[H]
    \centering
    \includegraphics[width=\textwidth]{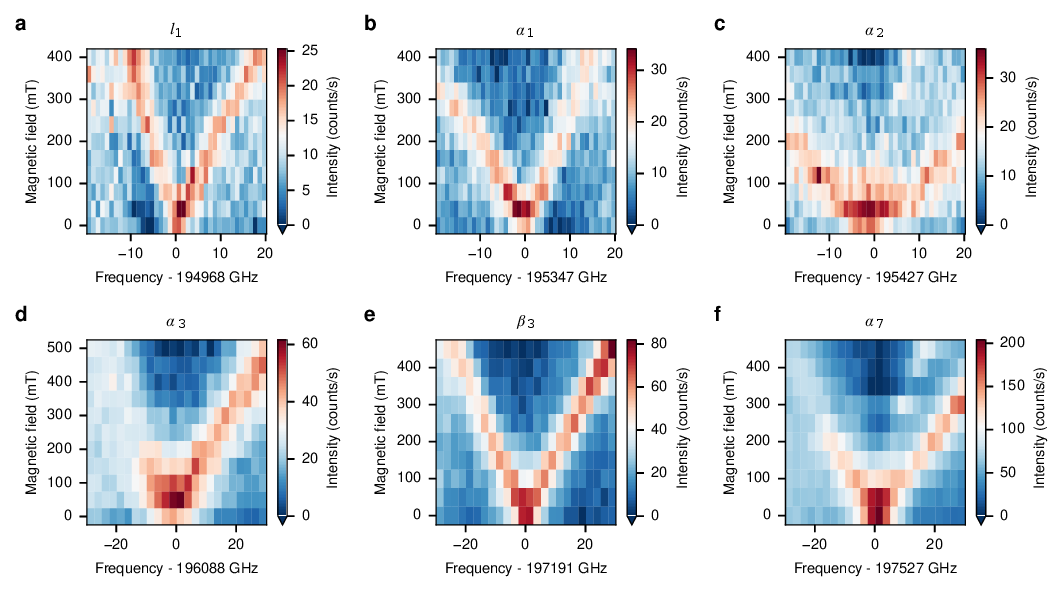}
    \caption{Zeeman splitting versus the applied magnetic field: \textbf{a} resonance $l_{1}$ at 194968 GHz exhibiting a $\Delta g$= (69.09 $\pm$ 1.08) GHz/T; \textbf{b} resonance $\alpha_{1}$ at 195347 GHz exhibiting a $\Delta g$ = (93.42 $\pm$ 3.12) GHz/T; \textbf{c} resonance $\alpha_{2}$ at 195427 GHz exhibiting a $\Delta g$ = (184.83 $\pm$ 9.54) GHz/T; \textbf{d} resonance $\alpha_{3}$ at 196088 GHz exhibiting a $\Delta g$ = (111.48 $\pm$ 2.69) GHz/T; \textbf{e} resonance $\beta_{2}$ at 197191 GHz exhibiting a $\Delta g$ = (116.56 $\pm$ 1.69) GHz/T; \textbf{f} resonance $\alpha_{7}$ at 197527 GHz exhibiting a $\Delta g$= (158.86 $\pm$ 3.73) GHz/T.}
    \label{fig3}
\end{figure}

Figure \ref{fig3} shows \ac{PLE} spectra of the brightest \erthr transitions measured at magnetic fields ranging from 0 to 400 mT applied in the plane of the SiC device layer.  \erthr optical transitions are known to split into multiple lines if the Zeeman splitting threshold exceeds \erthr inhomogeneous broadening~\cite{K_Thonke_1988, Sun_2008_magnetic, berkman2023observing, berkman2025long}. Both like-like and like-unlike \erthr \ac{PLE} transitions are typically observed in \ac{PLE} spectra~\cite{yin2013optical,PhysRevB.102.155309,berkman2023observing,holzapfel2024characterization,berkman2025long}. The Zeeman splitting of like-like (like-unlike) optical transitions is defined by the difference (sum) between Zeeman spitting of addressed \Ilow and \Ihigh energy levels with a strong reliance on the applied magnetic field direction and \erthr site symmetry \cite{dieke1970spectra, yang2022zeeman,holzapfel2024characterization}. Depending on the magnetic field orientation, the Zeeman splitting, $\Delta g$, of \Ilow and \Ihigh crystal field energy levels can vary from a few tens of GHz/T to $\sim$200 GHz/T~\cite{yang2022zeeman}, which translates to $\sim$0-50 GHz/T and $\sim$100-400 GHz/T splitting of like-like and like-unlike transitions in Si due to the misorientation of the \Ilow and \Ihigh g-tensors \cite{yin2013optical, yang2022zeeman,holzapfel2024characterization,berkman2025long}.

The \ac{PLE} lines of all observed \erthr sites in SiC split only into two lines, which indicates only a single possible \Ilow and \Ihigh g-tensor orientation for all observed \erthr transitions. Furthermore, only two observed Zeeman-split transitions could be explained by either strong selection rules for like-like and like-unlike \erthr transitions, i.e., allowed (not allowed) like-like (like-unlike) transitions, or the Zeeman splitting of the \Ihigh or the ground state energy levels below the inhomogeneous broadening. The latter translates into the splitting of much less than 16 GHz/T corresponding to $g_{inplane}<1.14$ for the $\alpha$ ground state since $\alpha$ \ac{PLE} lines have different Zeeman splitting. In this case, the observed Zeeman splitting of \ac{PLE} lines shown in Fig. \ref{fig3} would correspond to the Zeeman splitting of the energy levels from the \Ihigh manifold. Magnetic field rotation measurements \cite{yang2022zeeman,holzapfel2024characterization} would provide the necessary additional information to fully explain the observation of only two \ac{PLE} transitions.

For the lowest \erthr resonance $l_{1}$, which is shown in Fig. \ref{fig3}\textbf{a}, we identify a Zeeman splitting $\Delta g$ of (69.09 $\pm$ 1.08) GHz/T. The observed g-factors of $\alpha$-site related resonances ranged from $\Delta g$ of (93.42 $\pm$ 3.12) GHz/T to (184.83 $\pm$ 9.54) GHz/T, 
as shown in Fig. \ref{fig3}\textbf{b} - \textbf{d}, \textbf{f}. For the brightest $\beta$-site resonance, we find a $\Delta g$ of (116.56 $\pm$ 1.69) GHz/T (see Fig. \ref{fig3}\textbf{e}), considering the $\beta_{3}$ resonance.
Via a rotation of the direction from the B-field, the full g-tensor could be extracted, enabling a very close modeling of these defects with theoretical spin Hamiltonian equations and extracting \Ilow and \Ihigh g-tensor values \cite{yang2022zeeman,holzapfel2024characterization}.   


\begin{figure}[H]
    \centering
    \includegraphics[width=\textwidth]{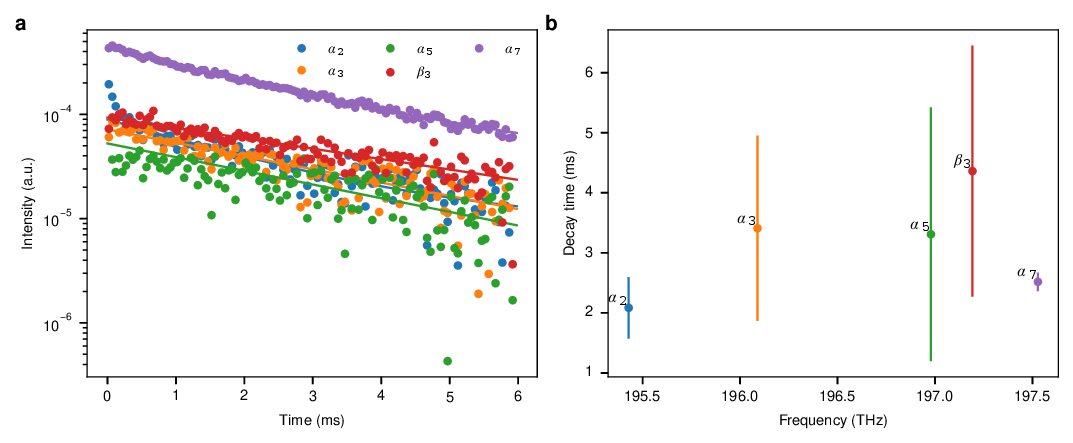}
    \caption{Optical lifetime: \textbf{a} Measured decay-transients over 6 ms in logarithmic scale for observable resonances $\alpha_{2}$, $\alpha_{3}$, $\alpha_{5}$, $\alpha_{7}$ and $\beta_{2}$ (dotted) and single exponential fits (solid lines) where optical lifetimes of (2.08 $\pm$ 0.3) ms, (3.41 $\pm$ 0.8) ms, (3.31 $\pm$ 1.1) ms, (2.52 $\pm$ 0.1) ms and (4.36 $\pm$ 1.1) ms were derived; \textbf{b} decay overview in dependence of observable resonance frequency where dots illustrate determined lifetimes. The error bars are from the fit uncertainties.}
    \label{fig4}
\end{figure}

Lastly, we investigated the optical lifetime from the brightest transitions of the identified $\alpha$ and $\beta$-\erthr-sites. This was determined by fitting the experimentally obtained transient PL data with a single exponential fit (see Methods). The $\alpha$-site related resonances exhibit a shorter lifetime with fitted values of (2.08 $\pm$ 0.3) ms, (3.41 $\pm$ 0.8) ms, (3.31 $\pm$ 1.1) ms and (2.52 $\pm$ 0.1) ms for $\alpha_{2}$, $\alpha_{3}$, $\alpha_{5}$ and $\alpha_{7}$ respectively. These values are the same within the standard errorbar since the optical decay is dominated by the optical relaxation from the two lowest energy levels within the \Ihigh manifold to the \Ilow energy levels. The $\beta_{2}$ transition exhibits a longer lifetime of (4.36 $\pm$ 1.1) ms, as shown in Fig. \ref{fig4}\textbf{a}, \textbf{b}. These long lifetimes are typically observed for \erthr-emitters embedded in oxide-materials \cite{phenicie2019narrow}, longer than commonly observed values in silicon \cite{berkman2023observing} which typically remain in the $\approx$ 1 ms regime at cryogenic temperatures. 


In conclusion, we have identified two Er$^{3+}$ sites ($\alpha$ and $\beta$) in 4H-SiCOI. Both exhibit remarkably narrow inhomogeneous and homogeneous linewidths, with the $\alpha$ site exhibiting values of $(6.44 \pm 0.55)$~GHz  and $(0.44 \pm 0.12)$~MHz, respectively, the narrowest values to date for any emitter in 4H-SiC. These values are competitive with Er$^{3+}$ in commercial un-optimised \ac{SOI}, where a similar implantation fluence yields inhomogeneous linewidths of 1-4~GHz and homogeneous linewidths of 30-100~MHz~\cite{weiss2021erbium}, and to rutile, where spectral diffusion reaches 267~MHz~\cite{DibosPurcell2022}. The sub-MHz homogeneous linewidths and reduced number of sites observed here highlight the exceptional optical quality of Er$^{3+}$ in 4H-SiCOI.

The measured optical lifetimes, (2.08 $\pm$ 0.3)~ms and (4.36 $\pm$ 1.1)~ms for sites $\alpha$ and $\beta$, respectively, place 4H-SiCOI between long-lived oxides~\cite{phenicie2019narrow} and the shorter sub-ms values observed in Si. This can be partially explained by a higher optical branching ratio of Er in SiC than the expected 0.2 branching ratio in Si~\cite{berkman2025long} as evident from one ($\alpha_{PL3}$) and two ($\beta_{PL1}$ and $\beta_{PL2}$) dominant \ac{PL} lines corresponding to the decay from the lowest \Ihigh crystal field levels of $\alpha$ and $\beta$ sites. The wider bandgap of SiC likely suppresses fast non-radiative channels, while the large $\Delta g$-tensor value of (116.56 $\pm$ 1.69)~GHz/T for the $\beta$ site exceeds that reported for Er$^{3+}$ in Si by nearly a factor of two~\cite{gritsch2023purcell}, illustrating the spin-photon interface potential of this system.


Looking forward, further improvements may be possible with lower nitrogen content epilayers and isotopically enriched 4H-SiCOI~\cite{steidl2025single}. Implantation annealing strategies may also be improved by incorporating the Er$^{3+}$ prior to 4H-SiCOI fabrication where the thermal budget is not limited by the SiO$_2$/Si handle wafer. The implantation depth of $\sim$250-300~nm demonstrated here is already compatible with integration into 4H-SiCOI nanophotonics, such as high-$Q$ ring resonators~\cite{CaiOctave22,bao2025tunable}. In addition, ensembles of Er$^{3+}$ emitters could be exploited with photon blockade~\cite{PhysRevResearch.4.033083}, while individual emitters preserve narrow optical linewidths even after near-surface integration~\cite{ourari2023indistinguishable, gritsch2022narrow}. Combined with the intrinsic advantages of SiC such as absence of two-photon absorption, $\chi^{(2)}$ and $\chi^{(3)}$ nonlinearities, low optical losses, broadband transparency, and full CMOS compatibility~\cite{wang2025scalable}, these findings establish Er$^{3+}$ in 4H-SiCOI as an appealing platform for scalable spin-photon interfaces and quantum memories.
\section*{Methods}
\subsection*{Sample preparation}
The 4H-SiCOI was fabricated with 630 nm thin 4H-SiC-layer on a 2 $\mu$m Silicon-dioxide ($\text{SiO}_\text{2}$)-layer, used for 4H-SiC wafer bonding to a 500 $\mu$m Si-wafer handle. The 4H-SiCOI was fabricated following wafer bonding and polishing processes as described in ref. \cite{CaiOctave22}. The original 4H-SiC used to form the 4H-SiCOI stack is a 500 $\mu$m thick, 100 mm epi-ready High Purity Semi-insulating (HPSI) 4H-SiC wafer substrate from CREE, grown on-axis, with resistivity of $\geq 10^5$ $\Omega\cdot\text{cm}$ and double sided chemical-mechanically polished (CMP).
Er-defects were created subsequently in the so formed 4H-SiCOI at the centre of the 4H-SiC thin layer via a two-step implantation process: 
(1) 1.5 MeV $1.0 \times 10^{12}$ $\text{Er}/\text{cm}^{2}$ and (2) 2 MeV $1.4 \times 10^{12}$ $\text{Er}/\text{cm}^{2}$, which result in a Gaussian distribution of \erthr ions with the FWHM of ~0.2 $\mu$m and the peak \erthr concentration of $1.2 \times 10^{17}$ $\text{Er}/\text{cm}^{3}$. The implantation was performed at 600 \textdegree{}C using $^{166}$\erthr predominantly and with 20\% contamination from $^{167}$\erthr. 
The sample was then subject to a 30 minutes 1000 \textdegree{}C annealing process in Argon-atmosphere. To avoid the occurrence of cracks within the SiC-layer during annealing, 100 $\mu$m wide trenches were etched into the SiC-layer of each sample utilizing photolithography.

\subsection*{\ac{PLE}}
The \ac{PLE}-study is carried out in a similar approach as presented in ref. \cite{bader2025photoluminescence}. We utilize a Leiden $^{4}\text{He}/^{3}\text{He}$ dry dilution refrigerator and \ac{CW} tunable excitation from 193.000 to 197.600 THz (1553 to 1517 nm) using the Pure Photonics PPCL550 low-noise tunable diode laser with the instantaneous linewidth of 10 kHz. The laser excitation is modulated using three \ac{AOM} to achieve $\sim$180 dB extinction ratio between 'on' and 'off' periods. We keep the excitation-pulse width,-power and -period consistent at 20 $\mu$s, 17 $\mu$W and 210 $\mu$s, respectively. We utilize the bias current from the employed superconducting single-photon detector (SSPD) to disable the detector during excitation events and establish a valid bias current 20 $\mu$s after excitation-extinction \cite{berkman2023observing}. For the wide \ac{PLE} spectral scans, the laser linewidth was broadened by generating a 60 MHz-wide spectral comb using a single iXblue MXAN-LN-10 amplitude \ac{EOM} driven by the \ac{RF} comb with evenly spaced lines supplied by Tektronix AWG5204. Lines with long optical coherence were identified by comparing spectra obtained with 60 MHz-broad optical excitation and the single frequency excitation.

\subsection*{\ac{PL} - site identification}
\ac{PLE} spectra of \erthr sites $\alpha$ and $\beta$ were identified by correlating \ac{PL} spectra obtained by optically exciting at the resonance frequencies of observed \ac{PLE} lines corresponding to transitions from the lowest energy level within the \Ilow manifold to \Ihigh crystal field-split energy levels. The resonance frequencies of \ac{PLE} lines and their \ac{FWHM} were extracted using single Gaussian fits of the inhomogeneous peaks observed in the \ac{PLE}-spectrum, with the fit error given as 95\% confidence interval. Following that, a \ac{TOF} with 30 GHz transmission bandwidth was implemented into the detection section of the experimental setup.  The laser was consecutively tuned in resonance with observed \ac{PLE} lines and the tunable filter was scanned within its full scanning range from 1527 to 1567 nm yielding a series of \ac{PL} spectra. Identical \ac{PL} resonance frequencies in measured \ac{PL} spectra indicated that the corresponding \ac{PLE} lines originate from the same \erthr site since the \ac{PL} spectrum of the same \erthr site has the same frequency components. Due to the fast phonon relaxation from the higher \Ihigh crystal-field energy levels, the \ac{PL} spectrum of the same \erthr site is dominated by allowed optical transitions from the lowest energy levels within the \Ihigh manifold to the crystal-field split energy levels within the \Ilow manifold.   

\subsection*{Spectral hole burning}
The homogeneous linewidths of \erthr ions was assessed using the spectral hole burning technique elaborated in an earlier work \cite{berkman2025long}. This technique is based on an optical frequency comb constituting $N_{exp}$ laser optical doublets with the optical detuning $\Delta f$ between optical fields within the optical doublet and the frequency separation $f_{comb}$ between the centre frequencies of the nearest optical doublets.  This technique allows significantly increasing \ac{PL} intensity and corresponding \ac{SNR} by repeating the spectral hole burning experiment over a large portion of the inhomogeneous line. The frequency separation between nearest optical doublets was kept much larger than the spectral hole linewidth at the used power.  The optical spectral comb was generated using a single iXblue MXAN-LN-10 amplitude \ac{EOM} driven by the \ac{RF} multitone signal supplied by Tektronix AWG5204 arbitrary waveform generator given by
\begin{equation}
    \frac{A}{N_{exp}}\sum_{i=1}^{N_{exp}}\cos\left(2\pi\left(f_{comb}\times\left\lceil\frac{i}{2}\right\rceil+\frac{(-1)^i\Delta f}{2}\right)t+\theta_{i-1}\right),
\end{equation}  
\noindent where $A$ is the scaling constant.
The Newman phase $\Theta_i = \pi i^2/N_{exp}$ provides the low crest factor waveform and the best power performance for the comb generation. Here, the 400 MHz \ac{RF} comb was used to generate 800 MHz optical comb.

\subsection*{Magnetic spectroscopy measurements} 
The Zeeman splitting measurement is conducted in a magnetic field ranging from 0 to 400 mT while simultaneously driving the determined inhomogeneous resonance with a 60-MHz broad optical excitation. Here, the tunable magnetic field is applied in the plane of the 4H-SiCOI device layer. The g-factor values were extracted from fits to the magnetic field-split \erthr transitions which included both the linear and quadratic Zeeman splitting terms.

\subsection*{Optical lifetime} 
Optical lifetimes are obtained by exciting specific PLE-resonances, centered at the previously determined inhomogeneous peaks, and recording the photon events for 6 ms, in a similar approach as reported in Ref. \cite{berkman2023observing, berkman2025long}. 
Following that, a single exponential fit is performed as 
\begin{equation}
    \tau_{\text{fit}} = \text{a} \cdot e^{-\frac{\text{x}}{\tau}},
\end{equation}
where the optical lifetime $\tau$ is derived. The error of the fit is given as the 95\% confidence interval. Due to fast phonon relaxation within the \Ihigh manifold, the measured lifetime does not depend on the \ac{PLE} resonance used for the \ac{PL} excitation \cite{berkman2025long} and was dominated by the optical decays from the lowest \Ihigh crystal field level to \Ilow crystal field levels.

\subsection*{The site $\alpha$ generation yield}
The yield of the \erthr site could be estimated from the \ac{PLE} measurements. The \ac{PLE} intensity of the $\alpha_1$ line at the 17 $\mu$W and $\gamma_{exc}=60$ MHz excitation bandwidth, or 0.13 $\mu$W within the $\alpha_1$ homogeneous linewidth is $I_{\alpha_1}=20$ counts/s (see Fig. \ref{fig1}). The expected power broadening of the homogeneous linewidth at this power is negligible (see Fig. \ref{fig2}$\mathbf{e}$) ensuring that only \erthr ions within the excitation bandwidth are excited. The collection efficiency can be approximated as $\eta_{fiber} = (1-\sqrt{1-(\text{NA}/n_{SiC})^2})/2$ where the numerical aperture of the used PM1550-XP fiber $\text{NA} = 0.125$ and the SiC refractive index $n_{SiC} = 2.6$. The resulting fiber collection efficiency is $\eta_{fiber}=0.0006$. The total detection efficiency is $\eta=0.0004$ accounting for the SSPD detection efficiency of $\eta_{SSPD}=0.65$ and the 90/10 beam splitter in the collection path. The total number of \erthr ions excited within the mode field diameter of $D_{M}=10.1$ $\mu$m could be estimated as $N_{Er}= \pi p \rho_{Er} D_{M}^2 h_{impl} \gamma_{exc}/(8 \gamma_{h})$. Here, we account for the radial dependence of the ion-optical mode field overlap defined by the normalized Gaussian distribution of the optical field, and the portion of the ions falling within the optical excitation bandwidth. Parameters $p<0.5$, $h_{impl}=0.2$ $\mu$m, and $\rho_{Er}=1.2 \times 10^{17}$ $\text{Er}/\text{cm}^{3}$ are the excitation probability at the maximum of the optical field, the FWHM of the Gaussian implantation profile within the SiC layer, and the peak implantation density. The maximum measured signal can be estimated as $I_{\alpha_1,max} = \eta N_{Er} t_{\alpha}^{-1}$ for $p=0.5$, or 600 counts/s. This gives a lower bound to the $\alpha$ site  yield of $I_{\alpha_1}/I_{\alpha_1,max} = 3\%$. Only two like-like transitions observed in the magneto-\ac{PLE} spectra ensure that this yield value holds for \erthr sites with well-defined spin-level splitting \cite{berkman2023observing,berkman2025long}.
The yield of the \erthr sites is generally not quoted for other CMOS-compatible platforms, such as Si. Recently, the \ac{PL} measurements of \erthr in a photonic cavity with a mode volume of around $0.8(\lambda/n)^3$ showed around 12 peaks \cite{gritsch2025optical}. The expected number of \erthr ions within this mode volume can be estimated as 750 accounting for the \erthr density of \text{$10^{14}$ cm$^{-3}$}. The resulting yield of \erthr in \ac{SOI} is 1.6\%. Sites with symmetry below the $T_d$ crystal symmetry will have multiple orientations that may be non-degenerate in the magnetic field \cite{berkman2023observing,berkman2025long} further reducing the \erthr site yield in Si with a well-defined spin properties below 0.8\% \cite{weiss2021erbium}.

\section*{Data availability}
Data underlying the results presented in this paper are not publicly available at this time but may be obtained from the authors upon reasonable request. Additional results can be found in the Supplementary Information.

\bibliography{sn-bibliography}


\begin{thebibliography}{72}
\ifx \bisbn   \undefined \def \bisbn  #1{ISBN #1}\fi
\ifx \binits  \undefined \def \binits#1{#1}\fi
\ifx \bauthor  \undefined \def \bauthor#1{#1}\fi
\ifx \batitle  \undefined \def \batitle#1{#1}\fi
\ifx \bjtitle  \undefined \def \bjtitle#1{#1}\fi
\ifx \bvolume  \undefined \def \bvolume#1{\textbf{#1}}\fi
\ifx \byear  \undefined \def \byear#1{#1}\fi
\ifx \bissue  \undefined \def \bissue#1{#1}\fi
\ifx \bfpage  \undefined \def \bfpage#1{#1}\fi
\ifx \blpage  \undefined \def \blpage #1{#1}\fi
\ifx \burl  \undefined \def \burl#1{\textsf{#1}}\fi
\ifx \doiurl  \undefined \def \doiurl#1{\url{https://doi.org/#1}}\fi
\ifx \betal  \undefined \def \betal{\textit{et al.}}\fi
\ifx \binstitute  \undefined \def \binstitute#1{#1}\fi
\ifx \binstitutionaled  \undefined \def \binstitutionaled#1{#1}\fi
\ifx \bctitle  \undefined \def \bctitle#1{#1}\fi
\ifx \beditor  \undefined \def \beditor#1{#1}\fi
\ifx \bpublisher  \undefined \def \bpublisher#1{#1}\fi
\ifx \bbtitle  \undefined \def \bbtitle#1{#1}\fi
\ifx \bedition  \undefined \def \bedition#1{#1}\fi
\ifx \bseriesno  \undefined \def \bseriesno#1{#1}\fi
\ifx \blocation  \undefined \def \blocation#1{#1}\fi
\ifx \bsertitle  \undefined \def \bsertitle#1{#1}\fi
\ifx \bsnm \undefined \def \bsnm#1{#1}\fi
\ifx \bsuffix \undefined \def \bsuffix#1{#1}\fi
\ifx \bparticle \undefined \def \bparticle#1{#1}\fi
\ifx \barticle \undefined \def \barticle#1{#1}\fi
\bibcommenthead
\ifx \bconfdate \undefined \def \bconfdate #1{#1}\fi
\ifx \botherref \undefined \def \botherref #1{#1}\fi
\ifx \url \undefined \def \url#1{\textsf{#1}}\fi
\ifx \bchapter \undefined \def \bchapter#1{#1}\fi
\ifx \bbook \undefined \def \bbook#1{#1}\fi
\ifx \bcomment \undefined \def \bcomment#1{#1}\fi
\ifx \oauthor \undefined \def \oauthor#1{#1}\fi
\ifx \citeauthoryear \undefined \def \citeauthoryear#1{#1}\fi
\ifx \endbibitem  \undefined \def \endbibitem {}\fi
\ifx \bconflocation  \undefined \def \bconflocation#1{#1}\fi
\ifx \arxivurl  \undefined \def \arxivurl#1{\textsf{#1}}\fi
\csname PreBibitemsHook\endcsname

\bibitem[\protect\citeauthoryear{Esparza et~al.}{2011}]{esparza2011high}
\begin{botherref}
\oauthor{\bsnm{Esparza}, \binits{L.R.}},
\oauthor{\bsnm{Childress}, \binits{L.}},
\oauthor{\bsnm{Bernien}, \binits{H.}},
\oauthor{\bsnm{Hensen}, \binits{B.}},
\oauthor{\bsnm{Alkemade}, \binits{P.}},
\oauthor{\bsnm{Hanson}, \binits{R.}}:
High-fidelity projective read-out of a solid-state spin quantum register.
Nature
\textbf{477}(7366)
(2011)
\end{botherref}
\endbibitem

\bibitem[\protect\citeauthoryear{Gao et~al.}{2015}]{gao2015coherent}
\begin{barticle}
\bauthor{\bsnm{Gao}, \binits{W.}},
\bauthor{\bsnm{Imamoglu}, \binits{A.}},
\bauthor{\bsnm{Bernien}, \binits{H.}},
\bauthor{\bsnm{Hanson}, \binits{R.}}:
\batitle{Coherent manipulation, measurement and entanglement of individual solid-state spins using optical fields}.
\bjtitle{Nature Photonics}
\bvolume{9}(\bissue{6}),
\bfpage{363}--\blpage{373}
(\byear{2015})
\end{barticle}
\endbibitem

\bibitem[\protect\citeauthoryear{Awschalom et~al.}{2018}]{awschalom2018quantum}
\begin{barticle}
\bauthor{\bsnm{Awschalom}, \binits{D.D.}},
\bauthor{\bsnm{Hanson}, \binits{R.}},
\bauthor{\bsnm{Wrachtrup}, \binits{J.}},
\bauthor{\bsnm{Zhou}, \binits{B.B.}}:
\batitle{Quantum technologies with optically interfaced solid-state spins}.
\bjtitle{Nature Photonics}
\bvolume{12}(\bissue{9}),
\bfpage{516}--\blpage{527}
(\byear{2018})
\end{barticle}
\endbibitem

\bibitem[\protect\citeauthoryear{Kindem et~al.}{2020}]{kindem2020control}
\begin{barticle}
\bauthor{\bsnm{Kindem}, \binits{J.M.}},
\bauthor{\bsnm{Ruskuc}, \binits{A.}},
\bauthor{\bsnm{Bartholomew}, \binits{J.G.}},
\bauthor{\bsnm{Rochman}, \binits{J.}},
\bauthor{\bsnm{Huan}, \binits{Y.Q.}},
\bauthor{\bsnm{Faraon}, \binits{A.}}:
\batitle{Control and single-shot readout of an ion embedded in a nanophotonic cavity}.
\bjtitle{Nature}
\bvolume{580}(\bissue{7802}),
\bfpage{201}--\blpage{204}
(\byear{2020})
\end{barticle}
\endbibitem

\bibitem[\protect\citeauthoryear{Kimble}{2008}]{kimble2008quantum}
\begin{barticle}
\bauthor{\bsnm{Kimble}, \binits{H.J.}}:
\batitle{The quantum internet}.
\bjtitle{Nature}
\bvolume{453}(\bissue{7198}),
\bfpage{1023}--\blpage{1030}
(\byear{2008})
\end{barticle}
\endbibitem

\bibitem[\protect\citeauthoryear{Wehner et~al.}{2018}]{wehner2018quantum}
\begin{barticle}
\bauthor{\bsnm{Wehner}, \binits{S.}},
\bauthor{\bsnm{Elkouss}, \binits{D.}},
\bauthor{\bsnm{Hanson}, \binits{R.}}:
\batitle{Quantum internet: A vision for the road ahead}.
\bjtitle{Science}
\bvolume{362}(\bissue{6412}),
\bfpage{9288}
(\byear{2018})
\end{barticle}
\endbibitem

\bibitem[\protect\citeauthoryear{Ran{\v{c}}i{\'c} et~al.}{2018}]{ranvcic2018coherence}
\begin{barticle}
\bauthor{\bsnm{Ran{\v{c}}i{\'c}}, \binits{M.}},
\bauthor{\bsnm{Hedges}, \binits{M.P.}},
\bauthor{\bsnm{Ahlefeldt}, \binits{R.L.}},
\bauthor{\bsnm{Sellars}, \binits{M.J.}}:
\batitle{Coherence time of over a second in a telecom-compatible quantum memory storage material}.
\bjtitle{Nature Physics}
\bvolume{14}(\bissue{1}),
\bfpage{50}--\blpage{54}
(\byear{2018})
\end{barticle}
\endbibitem

\bibitem[\protect\citeauthoryear{Ahn et~al.}{2024}]{JonghoonExtended2024}
\begin{barticle}
\bauthor{\bsnm{Ahn}, \binits{J.}},
\bauthor{\bsnm{Wicker}, \binits{C.}},
\bauthor{\bsnm{Bitner}, \binits{N.}},
\bauthor{\bsnm{Solomon}, \binits{M.T.}},
\bauthor{\bsnm{Tissot}, \binits{B.}},
\bauthor{\bsnm{Burkard}, \binits{G.}},
\bauthor{\bsnm{Dibos}, \binits{A.M.}},
\bauthor{\bsnm{Zhang}, \binits{J.}},
\bauthor{\bsnm{Heremans}, \binits{F.J.}},
\bauthor{\bsnm{Awschalom}, \binits{D.D.}}:
\batitle{Extended spin relaxation times of optically addressed vanadium defects in silicon carbide at telecommunication frequencies}.
\bjtitle{Phys. Rev. Appl.}
\bvolume{22},
\bfpage{044078}
(\byear{2024})
\end{barticle}
\endbibitem

\bibitem[\protect\citeauthoryear{Zhou et~al.}{2025}]{Zhou_quantum_networks2025}
\begin{barticle}
\bauthor{\bsnm{Zhou}, \binits{Y.}},
\bauthor{\bsnm{Tan}, \binits{J.}},
\bauthor{\bsnm{Hu}, \binits{H.}},
\bauthor{\bsnm{Hua}, \binits{S.}},
\bauthor{\bsnm{Jiang}, \binits{C.}},
\bauthor{\bsnm{Liang}, \binits{B.}},
\bauthor{\bsnm{Bao}, \binits{T.}},
\bauthor{\bsnm{Nie}, \binits{X.}},
\bauthor{\bsnm{Xiao}, \binits{S.}},
\bauthor{\bsnm{Lu}, \binits{D.}},
\bauthor{\bsnm{Wang}, \binits{J.}},
\bauthor{\bsnm{Song}, \binits{Q.}}:
\batitle{Silicon carbide: A promising platform for scalable quantum networks}.
\bjtitle{Applied Physics Reviews}
\bvolume{12}(\bissue{3}),
\bfpage{031301}
(\byear{2025})
\end{barticle}
\endbibitem

\bibitem[\protect\citeauthoryear{Gritsch et~al.}{2025}]{gritsch2025optical}
\begin{barticle}
\bauthor{\bsnm{Gritsch}, \binits{A.}},
\bauthor{\bsnm{Ulanowski}, \binits{A.}},
\bauthor{\bsnm{Pforr}, \binits{J.}},
\bauthor{\bsnm{Reiserer}, \binits{A.}}:
\batitle{Optical single-shot readout of spin qubits in silicon}.
\bjtitle{Nature Communications}
\bvolume{16}(\bissue{1}),
\bfpage{64}
(\byear{2025})
\end{barticle}
\endbibitem

\bibitem[\protect\citeauthoryear{Wolfowicz et~al.}{2021}]{wolfowicz2021quantum}
\begin{barticle}
\bauthor{\bsnm{Wolfowicz}, \binits{G.}},
\bauthor{\bsnm{Heremans}, \binits{F.J.}},
\bauthor{\bsnm{Anderson}, \binits{C.P.}},
\bauthor{\bsnm{Kanai}, \binits{S.}},
\bauthor{\bsnm{Seo}, \binits{H.}},
\bauthor{\bsnm{Gali}, \binits{A.}},
\bauthor{\bsnm{Galli}, \binits{G.}},
\bauthor{\bsnm{Awschalom}, \binits{D.D.}}:
\batitle{Quantum guidelines for solid-state spin defects}.
\bjtitle{Nature Reviews Materials}
\bvolume{6}(\bissue{10}),
\bfpage{906}--\blpage{925}
(\byear{2021})
\end{barticle}
\endbibitem

\bibitem[\protect\citeauthoryear{Pompili et~al.}{2021}]{pompili2021realization}
\begin{barticle}
\bauthor{\bsnm{Pompili}, \binits{M.}},
\bauthor{\bsnm{Hermans}, \binits{S.L.}},
\bauthor{\bsnm{Baier}, \binits{S.}},
\bauthor{\bsnm{Beukers}, \binits{H.K.}},
\bauthor{\bsnm{Humphreys}, \binits{P.C.}},
\bauthor{\bsnm{Schouten}, \binits{R.N.}},
\bauthor{\bsnm{Vermeulen}, \binits{R.F.}},
\bauthor{\bsnm{Tiggelman}, \binits{M.J.}},
\bauthor{\bsnm{Santos~Martins}, \binits{L.}},
\bauthor{\bsnm{Dirkse}, \binits{B.}}, \betal:
\batitle{Realization of a multinode quantum network of remote solid-state qubits}.
\bjtitle{Science}
\bvolume{372}(\bissue{6539}),
\bfpage{259}--\blpage{264}
(\byear{2021})
\end{barticle}
\endbibitem

\bibitem[\protect\citeauthoryear{Parker et~al.}{2024}]{parker2024diamond}
\begin{barticle}
\bauthor{\bsnm{Parker}, \binits{R.A.}},
\bauthor{\bsnm{Arjona~Mart{\'\i}nez}, \binits{J.}},
\bauthor{\bsnm{Chen}, \binits{K.C.}},
\bauthor{\bsnm{Stramma}, \binits{A.M.}},
\bauthor{\bsnm{Harris}, \binits{I.B.}},
\bauthor{\bsnm{Michaels}, \binits{C.P.}},
\bauthor{\bsnm{Trusheim}, \binits{M.E.}},
\bauthor{\bsnm{Hayhurst~Appel}, \binits{M.}},
\bauthor{\bsnm{Purser}, \binits{C.M.}},
\bauthor{\bsnm{Roth}, \binits{W.G.}}, \betal:
\batitle{A diamond nanophotonic interface with an optically accessible deterministic electronuclear spin register}.
\bjtitle{Nature Photonics}
\bvolume{18}(\bissue{2}),
\bfpage{156}--\blpage{161}
(\byear{2024})
\end{barticle}
\endbibitem

\bibitem[\protect\citeauthoryear{Redjem et~al.}{2020}]{redjem2020single}
\begin{barticle}
\bauthor{\bsnm{Redjem}, \binits{W.}},
\bauthor{\bsnm{Durand}, \binits{A.}},
\bauthor{\bsnm{Herzig}, \binits{T.}},
\bauthor{\bsnm{Benali}, \binits{A.}},
\bauthor{\bsnm{Pezzagna}, \binits{S.}},
\bauthor{\bsnm{Meijer}, \binits{J.}},
\bauthor{\bsnm{Kuznetsov}, \binits{A.Y.}},
\bauthor{\bsnm{Nguyen}, \binits{H.}},
\bauthor{\bsnm{Cueff}, \binits{S.}},
\bauthor{\bsnm{G{\'e}rard}, \binits{J.-M.}}, \betal:
\batitle{Single artificial atoms in silicon emitting at telecom wavelengths}.
\bjtitle{Nature Electronics}
\bvolume{3}(\bissue{12}),
\bfpage{738}--\blpage{743}
(\byear{2020})
\end{barticle}
\endbibitem

\bibitem[\protect\citeauthoryear{Stern et~al.}{2022}]{stern2022room}
\begin{barticle}
\bauthor{\bsnm{Stern}, \binits{H.L.}},
\bauthor{\bsnm{Gu}, \binits{Q.}},
\bauthor{\bsnm{Jarman}, \binits{J.}},
\bauthor{\bsnm{Eizagirre~Barker}, \binits{S.}},
\bauthor{\bsnm{Mendelson}, \binits{N.}},
\bauthor{\bsnm{Chugh}, \binits{D.}},
\bauthor{\bsnm{Schott}, \binits{S.}},
\bauthor{\bsnm{Tan}, \binits{H.H.}},
\bauthor{\bsnm{Sirringhaus}, \binits{H.}},
\bauthor{\bsnm{Aharonovich}, \binits{I.}}, \betal:
\batitle{Room-temperature optically detected magnetic resonance of single defects in hexagonal boron nitride}.
\bjtitle{Nature communications}
\bvolume{13}(\bissue{1}),
\bfpage{618}
(\byear{2022})
\end{barticle}
\endbibitem

\bibitem[\protect\citeauthoryear{Morioka et~al.}{2020}]{morioka2020spin}
\begin{barticle}
\bauthor{\bsnm{Morioka}, \binits{N.}},
\bauthor{\bsnm{Babin}, \binits{C.}},
\bauthor{\bsnm{Nagy}, \binits{R.}},
\bauthor{\bsnm{Gediz}, \binits{I.}},
\bauthor{\bsnm{Hesselmeier}, \binits{E.}},
\bauthor{\bsnm{Liu}, \binits{D.}},
\bauthor{\bsnm{Joliffe}, \binits{M.}},
\bauthor{\bsnm{Niethammer}, \binits{M.}},
\bauthor{\bsnm{Dasari}, \binits{D.}},
\bauthor{\bsnm{Vorobyov}, \binits{V.}}, \betal:
\batitle{Spin-controlled generation of indistinguishable and distinguishable photons from silicon vacancy centres in silicon carbide}.
\bjtitle{Nature communications}
\bvolume{11}(\bissue{1}),
\bfpage{2516}
(\byear{2020})
\end{barticle}
\endbibitem

\bibitem[\protect\citeauthoryear{Ourari et~al.}{2023}]{ourari2023indistinguishable}
\begin{barticle}
\bauthor{\bsnm{Ourari}, \binits{S.}},
\bauthor{\bsnm{Dusanowski}, \binits{{\L}.}},
\bauthor{\bsnm{Horvath}, \binits{S.P.}},
\bauthor{\bsnm{Uysal}, \binits{M.T.}},
\bauthor{\bsnm{Phenicie}, \binits{C.M.}},
\bauthor{\bsnm{Stevenson}, \binits{P.}},
\bauthor{\bsnm{Raha}, \binits{M.}},
\bauthor{\bsnm{Chen}, \binits{S.}},
\bauthor{\bsnm{Cava}, \binits{R.J.}},
\bauthor{\bsnm{Leon}, \binits{N.P.}}, \betal:
\batitle{Indistinguishable telecom band photons from a single er ion in the solid state}.
\bjtitle{Nature}
\bvolume{620}(\bissue{7976}),
\bfpage{977}--\blpage{981}
(\byear{2023})
\end{barticle}
\endbibitem

\bibitem[\protect\citeauthoryear{Knaut et~al.}{2024}]{knaut2024entanglement}
\begin{barticle}
\bauthor{\bsnm{Knaut}, \binits{C.M.}},
\bauthor{\bsnm{Suleymanzade}, \binits{A.}},
\bauthor{\bsnm{Wei}, \binits{Y.-C.}},
\bauthor{\bsnm{Assumpcao}, \binits{D.R.}},
\bauthor{\bsnm{Stas}, \binits{P.-J.}},
\bauthor{\bsnm{Huan}, \binits{Y.Q.}},
\bauthor{\bsnm{Machielse}, \binits{B.}},
\bauthor{\bsnm{Knall}, \binits{E.N.}},
\bauthor{\bsnm{Sutula}, \binits{M.}},
\bauthor{\bsnm{Baranes}, \binits{G.}}, \betal:
\batitle{Entanglement of nanophotonic quantum memory nodes in a telecom network}.
\bjtitle{Nature}
\bvolume{629}(\bissue{8012}),
\bfpage{573}--\blpage{578}
(\byear{2024})
\end{barticle}
\endbibitem

\bibitem[\protect\citeauthoryear{Son et~al.}{2020}]{son2020developing}
\begin{botherref}
\oauthor{\bsnm{Son}, \binits{N.T.}},
\oauthor{\bsnm{Anderson}, \binits{C.P.}},
\oauthor{\bsnm{Bourassa}, \binits{A.}},
\oauthor{\bsnm{Miao}, \binits{K.C.}},
\oauthor{\bsnm{Babin}, \binits{C.}},
\oauthor{\bsnm{Widmann}, \binits{M.}},
\oauthor{\bsnm{Niethammer}, \binits{M.}},
\oauthor{\bsnm{Ul~Hassan}, \binits{J.}},
\oauthor{\bsnm{Morioka}, \binits{N.}},
\oauthor{\bsnm{Ivanov}, \binits{I.G.}}, et al.:
Developing silicon carbide for quantum spintronics.
Applied Physics Letters
\textbf{116}(19)
(2020)
\end{botherref}
\endbibitem

\bibitem[\protect\citeauthoryear{Castelletto et~al.}{2022}]{castelletto2022silicon}
\begin{barticle}
\bauthor{\bsnm{Castelletto}, \binits{S.}},
\bauthor{\bsnm{Peruzzo}, \binits{A.}},
\bauthor{\bsnm{Bonato}, \binits{C.}},
\bauthor{\bsnm{Johnson}, \binits{B.C.}},
\bauthor{\bsnm{Radulaski}, \binits{M.}},
\bauthor{\bsnm{Ou}, \binits{H.}},
\bauthor{\bsnm{Kaiser}, \binits{F.}},
\bauthor{\bsnm{Wrachtrup}, \binits{J.}}:
\batitle{Silicon carbide photonics bridging quantum technology}.
\bjtitle{ACS Photonics}
\bvolume{9}(\bissue{5}),
\bfpage{1434}--\blpage{1457}
(\byear{2022})
\end{barticle}
\endbibitem

\bibitem[\protect\citeauthoryear{Widmann et~al.}{2015}]{widmann2015coherent}
\begin{barticle}
\bauthor{\bsnm{Widmann}, \binits{M.}},
\bauthor{\bsnm{Lee}, \binits{S.-Y.}},
\bauthor{\bsnm{Rendler}, \binits{T.}},
\bauthor{\bsnm{Son}, \binits{N.T.}},
\bauthor{\bsnm{Fedder}, \binits{H.}},
\bauthor{\bsnm{Paik}, \binits{S.}},
\bauthor{\bsnm{Yang}, \binits{L.-P.}},
\bauthor{\bsnm{Zhao}, \binits{N.}},
\bauthor{\bsnm{Yang}, \binits{S.}},
\bauthor{\bsnm{Booker}, \binits{I.}}, \betal:
\batitle{Coherent control of single spins in silicon carbide at room temperature}.
\bjtitle{Nature materials}
\bvolume{14}(\bissue{2}),
\bfpage{164}--\blpage{168}
(\byear{2015})
\end{barticle}
\endbibitem

\bibitem[\protect\citeauthoryear{Christle et~al.}{2015}]{christle2015isolated}
\begin{barticle}
\bauthor{\bsnm{Christle}, \binits{D.J.}},
\bauthor{\bsnm{Falk}, \binits{A.L.}},
\bauthor{\bsnm{Andrich}, \binits{P.}},
\bauthor{\bsnm{Klimov}, \binits{P.V.}},
\bauthor{\bsnm{Hassan}, \binits{J.U.}},
\bauthor{\bsnm{Son}, \binits{N.T.}},
\bauthor{\bsnm{Janz{\'e}n}, \binits{E.}},
\bauthor{\bsnm{Ohshima}, \binits{T.}},
\bauthor{\bsnm{Awschalom}, \binits{D.D.}}:
\batitle{Isolated electron spins in silicon carbide with millisecond coherence times}.
\bjtitle{Nature materials}
\bvolume{14}(\bissue{2}),
\bfpage{160}--\blpage{163}
(\byear{2015})
\end{barticle}
\endbibitem

\bibitem[\protect\citeauthoryear{Simin et~al.}{2017}]{simin2017locking}
\begin{barticle}
\bauthor{\bsnm{Simin}, \binits{D.}},
\bauthor{\bsnm{Kraus}, \binits{H.}},
\bauthor{\bsnm{Sperlich}, \binits{A.}},
\bauthor{\bsnm{Ohshima}, \binits{T.}},
\bauthor{\bsnm{Astakhov}, \binits{G.}},
\bauthor{\bsnm{Dyakonov}, \binits{V.}}:
\batitle{Locking of electron spin coherence above 20 ms in natural silicon carbide}.
\bjtitle{Physical Review B}
\bvolume{95}(\bissue{16}),
\bfpage{161201}
(\byear{2017})
\end{barticle}
\endbibitem

\bibitem[\protect\citeauthoryear{Nagy et~al.}{2019}]{nagy2019high}
\begin{barticle}
\bauthor{\bsnm{Nagy}, \binits{R.}},
\bauthor{\bsnm{Niethammer}, \binits{M.}},
\bauthor{\bsnm{Widmann}, \binits{M.}},
\bauthor{\bsnm{Chen}, \binits{Y.-C.}},
\bauthor{\bsnm{Udvarhelyi}, \binits{P.}},
\bauthor{\bsnm{Bonato}, \binits{C.}},
\bauthor{\bsnm{Hassan}, \binits{J.U.}},
\bauthor{\bsnm{Karhu}, \binits{R.}},
\bauthor{\bsnm{Ivanov}, \binits{I.G.}},
\bauthor{\bsnm{Son}, \binits{N.T.}}, \betal:
\batitle{High-fidelity spin and optical control of single silicon-vacancy centres in silicon carbide}.
\bjtitle{Nature communications}
\bvolume{10}(\bissue{1}),
\bfpage{1954}
(\byear{2019})
\end{barticle}
\endbibitem

\bibitem[\protect\citeauthoryear{Anderson et~al.}{2022}]{anderson2022five}
\begin{barticle}
\bauthor{\bsnm{Anderson}, \binits{C.P.}},
\bauthor{\bsnm{Glen}, \binits{E.O.}},
\bauthor{\bsnm{Zeledon}, \binits{C.}},
\bauthor{\bsnm{Bourassa}, \binits{A.}},
\bauthor{\bsnm{Jin}, \binits{Y.}},
\bauthor{\bsnm{Zhu}, \binits{Y.}},
\bauthor{\bsnm{Vorwerk}, \binits{C.}},
\bauthor{\bsnm{Crook}, \binits{A.L.}},
\bauthor{\bsnm{Abe}, \binits{H.}},
\bauthor{\bsnm{Ul-Hassan}, \binits{J.}}, \betal:
\batitle{Five-second coherence of a single spin with single-shot readout in silicon carbide}.
\bjtitle{Science advances}
\bvolume{8}(\bissue{5}),
\bfpage{5912}
(\byear{2022})
\end{barticle}
\endbibitem

\bibitem[\protect\citeauthoryear{Zeledon et~al.}{2025}]{zeledon2025minutelongquantumcoherenceenabled}
\begin{botherref}
\oauthor{\bsnm{Zeledon}, \binits{C.}},
\oauthor{\bsnm{Pingault}, \binits{B.}},
\oauthor{\bsnm{Marcks}, \binits{J.C.}},
\oauthor{\bsnm{Onizhuk}, \binits{M.}},
\oauthor{\bsnm{Tsaturyan}, \binits{Y.}},
\oauthor{\bsnm{Wang}, \binits{Y.-x.}},
\oauthor{\bsnm{Soloway}, \binits{B.S.}},
\oauthor{\bsnm{Abe}, \binits{H.}},
\oauthor{\bsnm{Ghezellou}, \binits{M.}},
\oauthor{\bsnm{Ul-Hassan}, \binits{J.}},
\oauthor{\bsnm{Ohshima}, \binits{T.}},
\oauthor{\bsnm{Son}, \binits{N.T.}},
\oauthor{\bsnm{Heremans}, \binits{F.J.}},
\oauthor{\bsnm{Galli}, \binits{G.}},
\oauthor{\bsnm{Anderson}, \binits{C.P.}},
\oauthor{\bsnm{Awschalom}, \binits{D.D.}}:
Minute-long quantum coherence enabled by electrical depletion of magnetic noise
(2025).
\url{https://arxiv.org/abs/2504.13164}
\end{botherref}
\endbibitem

\bibitem[\protect\citeauthoryear{Bourassa et~al.}{2020}]{bourassa2020entanglement}
\begin{barticle}
\bauthor{\bsnm{Bourassa}, \binits{A.}},
\bauthor{\bsnm{Anderson}, \binits{C.P.}},
\bauthor{\bsnm{Miao}, \binits{K.C.}},
\bauthor{\bsnm{Onizhuk}, \binits{M.}},
\bauthor{\bsnm{Ma}, \binits{H.}},
\bauthor{\bsnm{Crook}, \binits{A.L.}},
\bauthor{\bsnm{Abe}, \binits{H.}},
\bauthor{\bsnm{Ul-Hassan}, \binits{J.}},
\bauthor{\bsnm{Ohshima}, \binits{T.}},
\bauthor{\bsnm{Son}, \binits{N.T.}}, \betal:
\batitle{Entanglement and control of single nuclear spins in isotopically engineered silicon carbide}.
\bjtitle{Nature Materials}
\bvolume{19}(\bissue{12}),
\bfpage{1319}--\blpage{1325}
(\byear{2020})
\end{barticle}
\endbibitem

\bibitem[\protect\citeauthoryear{Fang et~al.}{2024}]{FangSpin_Photon2024}
\begin{barticle}
\bauthor{\bsnm{Fang}, \binits{R.-Z.}},
\bauthor{\bsnm{Lai}, \binits{X.-Y.}},
\bauthor{\bsnm{Li}, \binits{T.}},
\bauthor{\bsnm{Su}, \binits{R.-Z.}},
\bauthor{\bsnm{Lu}, \binits{B.-W.}},
\bauthor{\bsnm{Yang}, \binits{C.-W.}},
\bauthor{\bsnm{Liu}, \binits{R.-Z.}},
\bauthor{\bsnm{Qiao}, \binits{Y.-K.}},
\bauthor{\bsnm{Li}, \binits{C.}},
\bauthor{\bsnm{He}, \binits{Z.-G.}},
\bauthor{\bsnm{Huang}, \binits{J.}},
\bauthor{\bsnm{Li}, \binits{H.}},
\bauthor{\bsnm{You}, \binits{L.-X.}},
\bauthor{\bsnm{Huo}, \binits{Y.-H.}},
\bauthor{\bsnm{Bao}, \binits{X.-H.}},
\bauthor{\bsnm{Pan}, \binits{J.-W.}}:
\batitle{Experimental generation of spin-photon entanglement in silicon carbide}.
\bjtitle{Phys. Rev. Lett.}
\bvolume{132},
\bfpage{160801}
(\byear{2024})
\doiurl{10.1103/PhysRevLett.132.160801}
\end{barticle}
\endbibitem

\bibitem[\protect\citeauthoryear{Cilibrizzi et~al.}{2023}]{cilibrizzi2023ultra}
\begin{barticle}
\bauthor{\bsnm{Cilibrizzi}, \binits{P.}},
\bauthor{\bsnm{Arshad}, \binits{M.J.}},
\bauthor{\bsnm{Tissot}, \binits{B.}},
\bauthor{\bsnm{Son}, \binits{N.T.}},
\bauthor{\bsnm{Ivanov}, \binits{I.G.}},
\bauthor{\bsnm{Astner}, \binits{T.}},
\bauthor{\bsnm{Koller}, \binits{P.}},
\bauthor{\bsnm{Ghezellou}, \binits{M.}},
\bauthor{\bsnm{Ul-Hassan}, \binits{J.}},
\bauthor{\bsnm{White}, \binits{D.}}, \betal:
\batitle{Ultra-narrow inhomogeneous spectral distribution of telecom-wavelength vanadium centres in isotopically-enriched silicon carbide}.
\bjtitle{Nature Communications}
\bvolume{14}(\bissue{1}),
\bfpage{8448}
(\byear{2023})
\end{barticle}
\endbibitem

\bibitem[\protect\citeauthoryear{Nishikawa et~al.}{2025}]{nishikawa2025coherent}
\begin{barticle}
\bauthor{\bsnm{Nishikawa}, \binits{T.}},
\bauthor{\bsnm{Morioka}, \binits{N.}},
\bauthor{\bsnm{Abe}, \binits{H.}},
\bauthor{\bsnm{Murata}, \binits{K.}},
\bauthor{\bsnm{Okajima}, \binits{K.}},
\bauthor{\bsnm{Ohshima}, \binits{T.}},
\bauthor{\bsnm{Tsuchida}, \binits{H.}},
\bauthor{\bsnm{Mizuochi}, \binits{N.}}:
\batitle{Coherent photoelectrical readout of single spins in silicon carbide at room temperature}.
\bjtitle{Nature Communications}
\bvolume{16}(\bissue{1}),
\bfpage{3405}
(\byear{2025})
\end{barticle}
\endbibitem

\bibitem[\protect\citeauthoryear{Anderson et~al.}{2019}]{anderson2019electrical}
\begin{barticle}
\bauthor{\bsnm{Anderson}, \binits{C.P.}},
\bauthor{\bsnm{Bourassa}, \binits{A.}},
\bauthor{\bsnm{Miao}, \binits{K.C.}},
\bauthor{\bsnm{Wolfowicz}, \binits{G.}},
\bauthor{\bsnm{Mintun}, \binits{P.J.}},
\bauthor{\bsnm{Crook}, \binits{A.L.}},
\bauthor{\bsnm{Abe}, \binits{H.}},
\bauthor{\bsnm{Ul~Hassan}, \binits{J.}},
\bauthor{\bsnm{Son}, \binits{N.T.}},
\bauthor{\bsnm{Ohshima}, \binits{T.}}, \betal:
\batitle{Electrical and optical control of single spins integrated in scalable semiconductor devices}.
\bjtitle{Science}
\bvolume{366}(\bissue{6470}),
\bfpage{1225}--\blpage{1230}
(\byear{2019})
\end{barticle}
\endbibitem

\bibitem[\protect\citeauthoryear{Crook et~al.}{2020}]{Crook2020}
\begin{barticle}
\bauthor{\bsnm{Crook}, \binits{A.L.}},
\bauthor{\bsnm{Anderson}, \binits{C.P.}},
\bauthor{\bsnm{Miao}, \binits{K.C.}},
\bauthor{\bsnm{Bourassa}, \binits{A.}},
\bauthor{\bsnm{Lee}, \binits{H.}},
\bauthor{\bsnm{Bayliss}, \binits{S.L.}},
\bauthor{\bsnm{Bracher}, \binits{D.O.}},
\bauthor{\bsnm{Zhang}, \binits{X.}},
\bauthor{\bsnm{Abe}, \binits{H.}},
\bauthor{\bsnm{Ohshima}, \binits{T.}},
\bauthor{\bsnm{Hu}, \binits{E.L.}},
\bauthor{\bsnm{Awschalom}, \binits{D.D.}}:
\batitle{Purcell enhancement of a single silicon carbide color center with coherent spin control}.
\bjtitle{Nano Letters}
\bvolume{20}(\bissue{5}),
\bfpage{3427}--\blpage{3434}
(\byear{2020})
\end{barticle}
\endbibitem

\bibitem[\protect\citeauthoryear{Lukin et~al.}{2020a}]{lukin20204h}
\begin{barticle}
\bauthor{\bsnm{Lukin}, \binits{D.M.}},
\bauthor{\bsnm{Dory}, \binits{C.}},
\bauthor{\bsnm{Guidry}, \binits{M.A.}},
\bauthor{\bsnm{Yang}, \binits{K.Y.}},
\bauthor{\bsnm{Mishra}, \binits{S.D.}},
\bauthor{\bsnm{Trivedi}, \binits{R.}},
\bauthor{\bsnm{Radulaski}, \binits{M.}},
\bauthor{\bsnm{Sun}, \binits{S.}},
\bauthor{\bsnm{Vercruysse}, \binits{D.}},
\bauthor{\bsnm{Ahn}, \binits{G.H.}}, \betal:
\batitle{4h-silicon-carbide-on-insulator for integrated quantum and nonlinear photonics}.
\bjtitle{Nature Photonics}
\bvolume{14}(\bissue{5}),
\bfpage{330}--\blpage{334}
(\byear{2020})
\end{barticle}
\endbibitem

\bibitem[\protect\citeauthoryear{Lukin et~al.}{2020b}]{PRXQuantum.1.020102}
\begin{barticle}
\bauthor{\bsnm{Lukin}, \binits{D.M.}},
\bauthor{\bsnm{Guidry}, \binits{M.A.}},
\bauthor{\bsnm{Vučković}, \binits{J.}}:
\batitle{Integrated quantum photonics with silicon carbide: Challenges and prospects}.
\bjtitle{PRX Quantum}
\bvolume{1},
\bfpage{020102}
(\byear{2020})
\doiurl{10.1103/PRXQuantum.1.020102}
\end{barticle}
\endbibitem

\bibitem[\protect\citeauthoryear{Babin et~al.}{2022}]{babin2022fabrication}
\begin{barticle}
\bauthor{\bsnm{Babin}, \binits{C.}},
\bauthor{\bsnm{St{\"o}hr}, \binits{R.}},
\bauthor{\bsnm{Morioka}, \binits{N.}},
\bauthor{\bsnm{Linkewitz}, \binits{T.}},
\bauthor{\bsnm{Steidl}, \binits{T.}},
\bauthor{\bsnm{W{\"o}rnle}, \binits{R.}},
\bauthor{\bsnm{Liu}, \binits{D.}},
\bauthor{\bsnm{Hesselmeier}, \binits{E.}},
\bauthor{\bsnm{Vorobyov}, \binits{V.}},
\bauthor{\bsnm{Denisenko}, \binits{A.}}, \betal:
\batitle{Fabrication and nanophotonic waveguide integration of silicon carbide colour centres with preserved spin-optical coherence}.
\bjtitle{Nature materials}
\bvolume{21}(\bissue{1}),
\bfpage{67}--\blpage{73}
(\byear{2022})
\end{barticle}
\endbibitem

\bibitem[\protect\citeauthoryear{Day et~al.}{2023}]{day2023laser}
\begin{barticle}
\bauthor{\bsnm{Day}, \binits{A.M.}},
\bauthor{\bsnm{Dietz}, \binits{J.R.}},
\bauthor{\bsnm{Sutula}, \binits{M.}},
\bauthor{\bsnm{Yeh}, \binits{M.}},
\bauthor{\bsnm{Hu}, \binits{E.L.}}:
\batitle{Laser writing of spin defects in nanophotonic cavities}.
\bjtitle{Nature Materials}
\bvolume{22}(\bissue{6}),
\bfpage{696}--\blpage{702}
(\byear{2023})
\end{barticle}
\endbibitem

\bibitem[\protect\citeauthoryear{Cai et~al.}{2022}]{CaiOctave22}
\begin{barticle}
\bauthor{\bsnm{Cai}, \binits{L.}},
\bauthor{\bsnm{Li}, \binits{J.}},
\bauthor{\bsnm{Wang}, \binits{R.}},
\bauthor{\bsnm{Li}, \binits{Q.}}:
\batitle{Octave-spanning microcomb generation in 4h-silicon-carbide-on-insulator photonics platform}.
\bjtitle{Photonics Res.}
\bvolume{10}(\bissue{4}),
\bfpage{870}--\blpage{876}
(\byear{2022})
\end{barticle}
\endbibitem

\bibitem[\protect\citeauthoryear{Yang et~al.}{2023}]{yang2023inverse}
\begin{barticle}
\bauthor{\bsnm{Yang}, \binits{J.}},
\bauthor{\bsnm{Guidry}, \binits{M.A.}},
\bauthor{\bsnm{Lukin}, \binits{D.M.}},
\bauthor{\bsnm{Yang}, \binits{K.}},
\bauthor{\bsnm{Vu{\v{c}}kovi{\'c}}, \binits{J.}}:
\batitle{Inverse-designed silicon carbide quantum and nonlinear photonics}.
\bjtitle{Light: Science \& Applications}
\bvolume{12}(\bissue{1}),
\bfpage{201}
(\byear{2023})
\end{barticle}
\endbibitem

\bibitem[\protect\citeauthoryear{Bader et~al.}{2024}]{bader2024analysis}
\begin{barticle}
\bauthor{\bsnm{Bader}, \binits{J.}},
\bauthor{\bsnm{Arianfard}, \binits{H.}},
\bauthor{\bsnm{Peruzzo}, \binits{A.}},
\bauthor{\bsnm{Castelletto}, \binits{S.}}:
\batitle{Analysis, recent challenges and capabilities of spin-photon interfaces in silicon carbide-on-insulator}.
\bjtitle{npj Nanophotonics}
\bvolume{1}(\bissue{1}),
\bfpage{29}
(\byear{2024})
\end{barticle}
\endbibitem

\bibitem[\protect\citeauthoryear{Hu et~al.}{2024}]{hu2024room}
\begin{barticle}
\bauthor{\bsnm{Hu}, \binits{H.}},
\bauthor{\bsnm{Zhou}, \binits{Y.}},
\bauthor{\bsnm{Yi}, \binits{A.}},
\bauthor{\bsnm{Bao}, \binits{T.}},
\bauthor{\bsnm{Liu}, \binits{C.}},
\bauthor{\bsnm{Luo}, \binits{Q.}},
\bauthor{\bsnm{Zhang}, \binits{Y.}},
\bauthor{\bsnm{Wang}, \binits{Z.}},
\bauthor{\bsnm{Li}, \binits{Q.}},
\bauthor{\bsnm{Lu}, \binits{D.}}, \betal:
\batitle{Room-temperature waveguide integrated quantum register in a semiconductor photonic platform}.
\bjtitle{Nature Communications}
\bvolume{15}(\bissue{1}),
\bfpage{10256}
(\byear{2024})
\end{barticle}
\endbibitem

\bibitem[\protect\citeauthoryear{Lipton et~al.}{2025}]{lipton2025low}
\begin{barticle}
\bauthor{\bsnm{Lipton}, \binits{J.}},
\bauthor{\bsnm{Yurash}, \binits{B.}},
\bauthor{\bsnm{Sorensen}, \binits{A.}},
\bauthor{\bsnm{Vajo}, \binits{J.}},
\bauthor{\bsnm{Whiteley}, \binits{S.}},
\bauthor{\bsnm{Wang}, \binits{T.}},
\bauthor{\bsnm{Huang}, \binits{B.}},
\bauthor{\bsnm{Bai}, \binits{X.}},
\bauthor{\bsnm{Portales}, \binits{A.}},
\bauthor{\bsnm{Rubin}, \binits{S.}}, \betal:
\batitle{Low-loss nanophotonic devices with chip-level uniformity and integrated color centers in sic-on-insulator}.
\bjtitle{ACS Photonics}
\bvolume{12}(\bissue{5}),
\bfpage{2397}--\blpage{2405}
(\byear{2025})
\end{barticle}
\endbibitem

\bibitem[\protect\citeauthoryear{Dr{\'e}au et~al.}{2018}]{dreau2018quantum}
\begin{barticle}
\bauthor{\bsnm{Dr{\'e}au}, \binits{A.}},
\bauthor{\bsnm{Tchebotareva}, \binits{A.}},
\bauthor{\bsnm{Mahdaoui}, \binits{A.E.}},
\bauthor{\bsnm{Bonato}, \binits{C.}},
\bauthor{\bsnm{Hanson}, \binits{R.}}:
\batitle{Quantum frequency conversion of single photons from a nitrogen-vacancy center in diamond to telecommunication wavelengths}.
\bjtitle{Physical review applied}
\bvolume{9}(\bissue{6}),
\bfpage{064031}
(\byear{2018})
\end{barticle}
\endbibitem

\bibitem[\protect\citeauthoryear{Bader et~al.}{2025}]{bader2025photoluminescence}
\begin{botherref}
\oauthor{\bsnm{Bader}, \binits{J.}},
\oauthor{\bsnm{Lim}, \binits{S.Q.}},
\oauthor{\bsnm{Inam}, \binits{F.A.}},
\oauthor{\bsnm{Lyasota}, \binits{A.}},
\oauthor{\bsnm{Johnson}, \binits{B.C.}},
\oauthor{\bsnm{Peruzzo}, \binits{A.}},
\oauthor{\bsnm{McCallum}, \binits{J.C.}},
\oauthor{\bsnm{Li}, \binits{Q.}},
\oauthor{\bsnm{Rogge}, \binits{S.}},
\oauthor{\bsnm{Castelletto}, \binits{S.}}:
Photoluminescence properties of ion-implanted er3+ defects in 4h-sicoi for integrated quantum photonics.
ACS Applied Nano Materials
(2025)
\end{botherref}
\endbibitem

\bibitem[\protect\citeauthoryear{Babunts et~al.}{2000}]{babunts2000properties}
\begin{barticle}
\bauthor{\bsnm{Babunts}, \binits{R.}},
\bauthor{\bsnm{Vetrov}, \binits{V.}},
\bauthor{\bsnm{Il’in}, \binits{I.}},
\bauthor{\bsnm{Mokhov}, \binits{E.}},
\bauthor{\bsnm{Romanov}, \binits{N.}},
\bauthor{\bsnm{Khramtsov}, \binits{V.}},
\bauthor{\bsnm{Baranov}, \binits{P.}}:
\batitle{Properties of erbium luminescence in bulk crystals of silicon carbide}.
\bjtitle{Physics of the Solid State}
\bvolume{42},
\bfpage{829}--\blpage{835}
(\byear{2000})
\end{barticle}
\endbibitem

\bibitem[\protect\citeauthoryear{Ammerlaan and Pajot}{1996}]{choyke1997crystalfield}
\begin{bbook}
\bauthor{\bsnm{Ammerlaan}, \binits{C.A.J.}},
\bauthor{\bsnm{Pajot}, \binits{B.}}:
\bbtitle{Proceedings of International Conference on Shallow-Level Centers in Semiconductors},
\bedition{1}st edn.,
p. \bfpage{552}.
\bpublisher{World Scientific Publishing Company},
\blocation{Amsterdam, The Netherland}
(\byear{1996})
\end{bbook}
\endbibitem

\bibitem[\protect\citeauthoryear{Raha et~al.}{2020}]{raha2020optical}
\begin{barticle}
\bauthor{\bsnm{Raha}, \binits{M.}},
\bauthor{\bsnm{Chen}, \binits{S.}},
\bauthor{\bsnm{Phenicie}, \binits{C.M.}},
\bauthor{\bsnm{Ourari}, \binits{S.}},
\bauthor{\bsnm{Dibos}, \binits{A.M.}},
\bauthor{\bsnm{Thompson}, \binits{J.D.}}:
\batitle{Optical quantum nondemolition measurement of a single rare earth ion qubit}.
\bjtitle{Nature Communications}
\bvolume{11}(\bissue{1}),
\bfpage{1605}
(\byear{2020})
\end{barticle}
\endbibitem

\bibitem[\protect\citeauthoryear{Berkman et~al.}{2023}]{berkman2023observing}
\begin{barticle}
\bauthor{\bsnm{Berkman}, \binits{I.R.}},
\bauthor{\bsnm{Lyasota}, \binits{A.}},
\bauthor{\bsnm{De~Boo}, \binits{G.G.}},
\bauthor{\bsnm{Bartholomew}, \binits{J.G.}},
\bauthor{\bsnm{Johnson}, \binits{B.C.}},
\bauthor{\bsnm{McCallum}, \binits{J.C.}},
\bauthor{\bsnm{Xu}, \binits{B.-B.}},
\bauthor{\bsnm{Xie}, \binits{S.}},
\bauthor{\bsnm{Ahlefeldt}, \binits{R.L.}},
\bauthor{\bsnm{Sellars}, \binits{M.J.}}, \betal:
\batitle{Observing er 3+ sites in si with an in situ single-photon detector}.
\bjtitle{Physical Review Applied}
\bvolume{19}(\bissue{1}),
\bfpage{014037}
(\byear{2023})
\end{barticle}
\endbibitem

\bibitem[\protect\citeauthoryear{Berkman et~al.}{2025}]{berkman2025long}
\begin{barticle}
\bauthor{\bsnm{Berkman}, \binits{I.R.}},
\bauthor{\bsnm{Lyasota}, \binits{A.}},
\bauthor{\bsnm{Boo}, \binits{G.G.}},
\bauthor{\bsnm{Bartholomew}, \binits{J.G.}},
\bauthor{\bsnm{Lim}, \binits{S.Q.}},
\bauthor{\bsnm{Johnson}, \binits{B.C.}},
\bauthor{\bsnm{McCallum}, \binits{J.C.}},
\bauthor{\bsnm{Xu}, \binits{B.-B.}},
\bauthor{\bsnm{Xie}, \binits{S.}},
\bauthor{\bsnm{Abrosimov}, \binits{N.V.}}, \betal:
\batitle{Long optical and electron spin coherence times for erbium ions in silicon}.
\bjtitle{npj Quantum Information}
\bvolume{11}(\bissue{1}),
\bfpage{66}
(\byear{2025})
\end{barticle}
\endbibitem

\bibitem[\protect\citeauthoryear{Weiss et~al.}{2021}]{weiss2021erbium}
\begin{barticle}
\bauthor{\bsnm{Weiss}, \binits{L.}},
\bauthor{\bsnm{Gritsch}, \binits{A.}},
\bauthor{\bsnm{Merkel}, \binits{B.}},
\bauthor{\bsnm{Reiserer}, \binits{A.}}:
\batitle{Erbium dopants in nanophotonic silicon waveguides}.
\bjtitle{Optica}
\bvolume{8}(\bissue{1}),
\bfpage{40}--\blpage{41}
(\byear{2021})
\end{barticle}
\endbibitem

\bibitem[\protect\citeauthoryear{Gritsch et~al.}{2022}]{gritsch2022narrow}
\begin{barticle}
\bauthor{\bsnm{Gritsch}, \binits{A.}},
\bauthor{\bsnm{Weiss}, \binits{L.}},
\bauthor{\bsnm{Frueh}, \binits{J.}},
\bauthor{\bsnm{Rinner}, \binits{S.}},
\bauthor{\bsnm{Reiserer}, \binits{A.}}:
\batitle{Narrow optical transitions in erbium-implanted silicon waveguides}.
\bjtitle{Physical Review X}
\bvolume{12}(\bissue{4}),
\bfpage{041009}
(\byear{2022})
\end{barticle}
\endbibitem

\bibitem[\protect\citeauthoryear{Liu and Jacquier}{2005}]{liu2006spectroscopic}
\begin{bbook}
\bauthor{\bsnm{Liu}, \binits{G.}},
\bauthor{\bsnm{Jacquier}, \binits{B.}}:
\bbtitle{Spectroscopic Properties of Rare Earths in Optical Materials}
vol. \bseriesno{83},
\bedition{1}st edn.,
p. \bfpage{550}.
\bpublisher{Springer},
\blocation{Berlin, Germany}
(\byear{2005})
\end{bbook}
\endbibitem

\bibitem[\protect\citeauthoryear{Szabo}{1975}]{szabo1975observation}
\begin{barticle}
\bauthor{\bsnm{Szabo}, \binits{A.}}:
\batitle{Observation of hole burning and cross relaxation effects in ruby}.
\bjtitle{Physical Review B}
\bvolume{11}(\bissue{11}),
\bfpage{4512}
(\byear{1975})
\end{barticle}
\endbibitem

\bibitem[\protect\citeauthoryear{Voelker}{1989}]{volker1989hole}
\begin{barticle}
\bauthor{\bsnm{Voelker}, \binits{S.}}:
\batitle{Hole-burning spectroscopy}.
\bjtitle{Annual Review of Physical Chemistry}
\bvolume{40}(\bissue{1}),
\bfpage{499}--\blpage{530}
(\byear{1989})
\end{barticle}
\endbibitem

\bibitem[\protect\citeauthoryear{Nagy et~al.}{2021}]{nagy2021narrow}
\begin{botherref}
\oauthor{\bsnm{Nagy}, \binits{R.}},
\oauthor{\bsnm{Dasari}, \binits{D.B.R.}},
\oauthor{\bsnm{Babin}, \binits{C.}},
\oauthor{\bsnm{Liu}, \binits{D.}},
\oauthor{\bsnm{Vorobyov}, \binits{V.}},
\oauthor{\bsnm{Niethammer}, \binits{M.}},
\oauthor{\bsnm{Widmann}, \binits{M.}},
\oauthor{\bsnm{Linkewitz}, \binits{T.}},
\oauthor{\bsnm{Gediz}, \binits{I.}},
\oauthor{\bsnm{St{\"o}hr}, \binits{R.}}, et al.:
Narrow inhomogeneous distribution of spin-active emitters in silicon carbide.
Applied Physics Letters
\textbf{118}(14)
(2021)
\end{botherref}
\endbibitem

\bibitem[\protect\citeauthoryear{Heiler et~al.}{2024}]{heiler2024spectral}
\begin{barticle}
\bauthor{\bsnm{Heiler}, \binits{J.}},
\bauthor{\bsnm{K{\"o}rber}, \binits{J.}},
\bauthor{\bsnm{Hesselmeier}, \binits{E.}},
\bauthor{\bsnm{Kuna}, \binits{P.}},
\bauthor{\bsnm{St{\"o}hr}, \binits{R.}},
\bauthor{\bsnm{Fuchs}, \binits{P.}},
\bauthor{\bsnm{Ghezellou}, \binits{M.}},
\bauthor{\bsnm{Ul-Hassan}, \binits{J.}},
\bauthor{\bsnm{Knolle}, \binits{W.}},
\bauthor{\bsnm{Becher}, \binits{C.}}, \betal:
\batitle{Spectral stability of v2 centres in sub-micron 4h-sic membranes}.
\bjtitle{npj Quantum Materials}
\bvolume{9}(\bissue{1}),
\bfpage{34}
(\byear{2024})
\end{barticle}
\endbibitem

\bibitem[\protect\citeauthoryear{He et~al.}{2024}]{he2024robust}
\begin{barticle}
\bauthor{\bsnm{He}, \binits{Z.-X.}},
\bauthor{\bsnm{Zhou}, \binits{J.-Y.}},
\bauthor{\bsnm{Li}, \binits{Q.}},
\bauthor{\bsnm{Lin}, \binits{W.-X.}},
\bauthor{\bsnm{Liang}, \binits{R.-J.}},
\bauthor{\bsnm{Wang}, \binits{J.-F.}},
\bauthor{\bsnm{Wen}, \binits{X.-L.}},
\bauthor{\bsnm{Hao}, \binits{Z.-H.}},
\bauthor{\bsnm{Liu}, \binits{W.}},
\bauthor{\bsnm{Ren}, \binits{S.}}, \betal:
\batitle{Robust single modified divacancy color centers in 4h-sic under resonant excitation}.
\bjtitle{Nature Communications}
\bvolume{15}(\bissue{1}),
\bfpage{10146}
(\byear{2024})
\end{barticle}
\endbibitem

\bibitem[\protect\citeauthoryear{Wolfowicz et~al.}{2020}]{wolfowicz2020vanadium}
\begin{barticle}
\bauthor{\bsnm{Wolfowicz}, \binits{G.}},
\bauthor{\bsnm{Anderson}, \binits{C.P.}},
\bauthor{\bsnm{Diler}, \binits{B.}},
\bauthor{\bsnm{Poluektov}, \binits{O.G.}},
\bauthor{\bsnm{Heremans}, \binits{F.J.}},
\bauthor{\bsnm{Awschalom}, \binits{D.D.}}:
\batitle{Vanadium spin qubits as telecom quantum emitters in silicon carbide}.
\bjtitle{Science advances}
\bvolume{6}(\bissue{18}),
\bfpage{1192}
(\byear{2020})
\end{barticle}
\endbibitem

\bibitem[\protect\citeauthoryear{Higginbottom et~al.}{2022}]{higginbottom2022optical}
\begin{barticle}
\bauthor{\bsnm{Higginbottom}, \binits{D.B.}},
\bauthor{\bsnm{Kurkjian}, \binits{A.T.}},
\bauthor{\bsnm{Chartrand}, \binits{C.}},
\bauthor{\bsnm{Kazemi}, \binits{M.}},
\bauthor{\bsnm{Brunelle}, \binits{N.A.}},
\bauthor{\bsnm{MacQuarrie}, \binits{E.R.}},
\bauthor{\bsnm{Klein}, \binits{J.R.}},
\bauthor{\bsnm{Lee-Hone}, \binits{N.R.}},
\bauthor{\bsnm{Stacho}, \binits{J.}},
\bauthor{\bsnm{Ruether}, \binits{M.}}, \betal:
\batitle{Optical observation of single spins in silicon}.
\bjtitle{Nature}
\bvolume{607}(\bissue{7918}),
\bfpage{266}--\blpage{270}
(\byear{2022})
\end{barticle}
\endbibitem

\bibitem[\protect\citeauthoryear{Steidl et~al.}{2025}]{steidl2025single}
\begin{barticle}
\bauthor{\bsnm{Steidl}, \binits{T.}},
\bauthor{\bsnm{Kuna}, \binits{P.}},
\bauthor{\bsnm{Hesselmeier-Huettmann}, \binits{E.}},
\bauthor{\bsnm{Liu}, \binits{D.}},
\bauthor{\bsnm{Stoehr}, \binits{R.}},
\bauthor{\bsnm{Knolle}, \binits{W.}},
\bauthor{\bsnm{Ghezellou}, \binits{M.}},
\bauthor{\bsnm{Ul-Hassan}, \binits{J.}},
\bauthor{\bsnm{Schober}, \binits{M.}},
\bauthor{\bsnm{Bockstedte}, \binits{M.}}, \betal:
\batitle{Single v2 defect in 4h silicon carbide schottky diode at low temperature}.
\bjtitle{Nature Communications}
\bvolume{16}(\bissue{1}),
\bfpage{1}--\blpage{7}
(\byear{2025})
\end{barticle}
\endbibitem

\bibitem[\protect\citeauthoryear{Thonke et~al.}{1988}]{K_Thonke_1988}
\begin{barticle}
\bauthor{\bsnm{Thonke}, \binits{K.}},
\bauthor{\bsnm{Hermann}, \binits{H.U.}},
\bauthor{\bsnm{Schneider}, \binits{J.}}:
\batitle{A zeeman study of the 1.54$\mu$m transition in molecular beam epitaxial gaas:er}.
\bjtitle{Journal of Physics C: Solid State Physics}
\bvolume{21}(\bissue{34}),
\bfpage{5881}
(\byear{1988})
\end{barticle}
\endbibitem

\bibitem[\protect\citeauthoryear{Sun et~al.}{2008}]{Sun_2008_magnetic}
\begin{barticle}
\bauthor{\bsnm{Sun}, \binits{Y.}},
\bauthor{\bsnm{B\"ottger}, \binits{T.}},
\bauthor{\bsnm{Thiel}, \binits{C.W.}},
\bauthor{\bsnm{Cone}, \binits{R.L.}}:
\batitle{Magnetic $g$ tensors for the $^{4}\mathrm{I}_{15/2}$ and $^{4}\mathrm{I}_{13/2}$ states of ${\mathrm{er}}^{3+}:{\mathrm{y}}_{2}\mathrm{Si}{\mathrm{o}}_{5}$}.
\bjtitle{Phys. Rev. B}
\bvolume{77},
\bfpage{085124}
(\byear{2008})
\end{barticle}
\endbibitem

\bibitem[\protect\citeauthoryear{Yin et~al.}{2013}]{yin2013optical}
\begin{barticle}
\bauthor{\bsnm{Yin}, \binits{C.}},
\bauthor{\bsnm{Rančić}, \binits{M.}},
\bauthor{\bsnm{De~Boo}, \binits{G.G.}},
\bauthor{\bsnm{Stavrias}, \binits{N.}},
\bauthor{\bsnm{McCallum}, \binits{J.C.}},
\bauthor{\bsnm{Sellars}, \binits{M.J.}},
\bauthor{\bsnm{Rogge}, \binits{S.}}:
\batitle{Optical addressing of an individual erbium ion in silicon}.
\bjtitle{Nature}
\bvolume{497}(\bissue{7447}),
\bfpage{91}--\blpage{94}
(\byear{2013})
\end{barticle}
\endbibitem

\bibitem[\protect\citeauthoryear{de~Boo et~al.}{2020}]{PhysRevB.102.155309}
\begin{barticle}
\bauthor{\bsnm{Boo}, \binits{G.G.}},
\bauthor{\bsnm{Yin}, \binits{C.}},
\bauthor{\bsnm{Rančić}, \binits{M.}},
\bauthor{\bsnm{Johnson}, \binits{B.C.}},
\bauthor{\bsnm{McCallum}, \binits{J.C.}},
\bauthor{\bsnm{Sellars}, \binits{M.J.}},
\bauthor{\bsnm{Rogge}, \binits{S.}}:
\batitle{High-resolution spectroscopy of individual erbium ions in strong magnetic fields}.
\bjtitle{Phys. Rev. B}
\bvolume{102},
\bfpage{155309}
(\byear{2020})
\doiurl{10.1103/PhysRevB.102.155309}
\end{barticle}
\endbibitem

\bibitem[\protect\citeauthoryear{Holzaepfel et~al.}{2024}]{holzapfel2024characterization}
\begin{botherref}
\oauthor{\bsnm{Holzaepfel}, \binits{A.}},
\oauthor{\bsnm{Rinner}, \binits{S.}},
\oauthor{\bsnm{Sandholzer}, \binits{K.}},
\oauthor{\bsnm{Gritsch}, \binits{A.}},
\oauthor{\bsnm{Chaneliere}, \binits{T.}},
\oauthor{\bsnm{Reiserer}, \binits{A.}}:
Characterization of the spin and crystal field hamiltonian of erbium dopants in silicon.
Advanced Quantum Technologies,
2400342
(2024)
\end{botherref}
\endbibitem

\bibitem[\protect\citeauthoryear{Dieke and Satten}{1970}]{dieke1970spectra}
\begin{barticle}
\bauthor{\bsnm{Dieke}, \binits{G.H.}},
\bauthor{\bsnm{Satten}, \binits{R.A.}}:
\batitle{Spectra and energy levels of rare earth ions in crystals}.
\bjtitle{American Journal of Physics}
\bvolume{38}(\bissue{3}),
\bfpage{399}--\blpage{400}
(\byear{1970})
\end{barticle}
\endbibitem

\bibitem[\protect\citeauthoryear{Yang et~al.}{2022}]{yang2022zeeman}
\begin{barticle}
\bauthor{\bsnm{Yang}, \binits{J.}},
\bauthor{\bsnm{Fan}, \binits{W.}},
\bauthor{\bsnm{Zhang}, \binits{Y.}},
\bauthor{\bsnm{Duan}, \binits{C.}},
\bauthor{\bsnm{De~Boo}, \binits{G.G.}},
\bauthor{\bsnm{Ahlefeldt}, \binits{R.L.}},
\bauthor{\bsnm{Longdell}, \binits{J.J.}},
\bauthor{\bsnm{Johnson}, \binits{B.C.}},
\bauthor{\bsnm{McCallum}, \binits{J.C.}},
\bauthor{\bsnm{Sellars}, \binits{M.J.}}, \betal:
\batitle{Zeeman and hyperfine interactions of a single er 3+ 167 ion in si}.
\bjtitle{Physical Review B}
\bvolume{105}(\bissue{23}),
\bfpage{235306}
(\byear{2022})
\end{barticle}
\endbibitem

\bibitem[\protect\citeauthoryear{Phenicie et~al.}{2019}]{phenicie2019narrow}
\begin{barticle}
\bauthor{\bsnm{Phenicie}, \binits{C.M.}},
\bauthor{\bsnm{Stevenson}, \binits{P.}},
\bauthor{\bsnm{Welinski}, \binits{S.}},
\bauthor{\bsnm{Rose}, \binits{B.C.}},
\bauthor{\bsnm{Asfaw}, \binits{A.T.}},
\bauthor{\bsnm{Cava}, \binits{R.J.}},
\bauthor{\bsnm{Lyon}, \binits{S.A.}},
\bauthor{\bsnm{De~Leon}, \binits{N.P.}},
\bauthor{\bsnm{Thompson}, \binits{J.D.}}:
\batitle{Narrow optical line widths in erbium implanted in tio2}.
\bjtitle{Nano letters}
\bvolume{19}(\bissue{12}),
\bfpage{8928}--\blpage{8933}
(\byear{2019})
\end{barticle}
\endbibitem

\bibitem[\protect\citeauthoryear{Dibos et~al.}{2022}]{DibosPurcell2022}
\begin{barticle}
\bauthor{\bsnm{Dibos}, \binits{A.M.}},
\bauthor{\bsnm{Solomon}, \binits{M.T.}},
\bauthor{\bsnm{Sullivan}, \binits{S.E.}},
\bauthor{\bsnm{Singh}, \binits{M.K.}},
\bauthor{\bsnm{Sautter}, \binits{K.E.}},
\bauthor{\bsnm{Horn}, \binits{C.P.}},
\bauthor{\bsnm{Grant}, \binits{G.D.}},
\bauthor{\bsnm{Lin}, \binits{Y.}},
\bauthor{\bsnm{Wen}, \binits{J.}},
\bauthor{\bsnm{Heremans}, \binits{F.J.}},
\bauthor{\bsnm{Guha}, \binits{S.}},
\bauthor{\bsnm{Awschalom}, \binits{D.D.}}:
\batitle{Purcell enhancement of erbium ions in tio2 on silicon nanocavities}.
\bjtitle{Nano Letters}
\bvolume{22}(\bissue{16}),
\bfpage{6530}--\blpage{6536}
(\byear{2022})
\doiurl{10.1021/acs.nanolett.2c01561}
{\href{https://arxiv.org/abs/https://doi.org/10.1021/acs.nanolett.2c01561}{{https://doi.org/10.1021/acs.nanolett.2c01561}}}.
\bcomment{PMID: 35939762}
\end{barticle}
\endbibitem

\bibitem[\protect\citeauthoryear{Gritsch et~al.}{2023}]{gritsch2023purcell}
\begin{barticle}
\bauthor{\bsnm{Gritsch}, \binits{A.}},
\bauthor{\bsnm{Ulanowski}, \binits{A.}},
\bauthor{\bsnm{Reiserer}, \binits{A.}}:
\batitle{Purcell enhancement of single-photon emitters in silicon}.
\bjtitle{Optica}
\bvolume{10}(\bissue{6}),
\bfpage{783}--\blpage{789}
(\byear{2023})
\end{barticle}
\endbibitem

\bibitem[\protect\citeauthoryear{Bao et~al.}{2025}]{bao2025tunable}
\begin{botherref}
\oauthor{\bsnm{Bao}, \binits{T.}},
\oauthor{\bsnm{Luo}, \binits{Q.}},
\oauthor{\bsnm{Yi}, \binits{A.}},
\oauthor{\bsnm{Liang}, \binits{B.}},
\oauthor{\bsnm{Zhang}, \binits{Y.}},
\oauthor{\bsnm{Hu}, \binits{H.-B.}},
\oauthor{\bsnm{Lai}, \binits{S.}},
\oauthor{\bsnm{Liu}, \binits{Z.}},
\oauthor{\bsnm{Xiao}, \binits{S.}},
\oauthor{\bsnm{Ou}, \binits{X.}}, et al.:
Tunable cavity coupling to spin defects in a 4h-silicon-carbide-on-insulator platform.
ACS Photonics
(2025)
\end{botherref}
\endbibitem

\bibitem[\protect\citeauthoryear{Chen et~al.}{2022}]{PhysRevResearch.4.033083}
\begin{barticle}
\bauthor{\bsnm{Chen}, \binits{M.}},
\bauthor{\bsnm{Tang}, \binits{J.}},
\bauthor{\bsnm{Tang}, \binits{L.}},
\bauthor{\bsnm{Wu}, \binits{H.}},
\bauthor{\bsnm{Xia}, \binits{K.}}:
\batitle{Photon blockade and single-photon generation with multiple quantum emitters}.
\bjtitle{Phys. Rev. Res.}
\bvolume{4},
\bfpage{033083}
(\byear{2022})
\doiurl{10.1103/PhysRevResearch.4.033083}
\end{barticle}
\endbibitem

\bibitem[\protect\citeauthoryear{Wang et~al.}{2025}]{wang2025scalable}
\begin{botherref}
\oauthor{\bsnm{Wang}, \binits{H.}},
\oauthor{\bsnm{Ralph}, \binits{T.C.}},
\oauthor{\bsnm{Renema}, \binits{J.J.}},
\oauthor{\bsnm{Lu}, \binits{C.-Y.}},
\oauthor{\bsnm{Pan}, \binits{J.-W.}}:
Scalable photonic quantum technologies.
Nature Materials,
1--15
(2025)
\end{botherref}
\endbibitem

\end{thebibliography}
\section*{Acknowledgments}
J.McC. acknowledges the Australian Government Australian Research Council under the Centre of Excellence scheme (No: CE170100012). 
Q.L. is supported by the National Science Foundation of the United States of America under Grant No. 2240420.
A.L. and S.R. are supported by the ARC Centre of Excellence for Quantum Computation and Communication Technology (Grant CE170100012) and the Discovery Project (Grant DP210101784).

We acknowledge the use of the NCRIS Heavy Ion Accelerator platform (HIA) for access and support to the ion implantation equipment at the Australian National University.

This work was performed in part at the RMIT Micro Nano Research Facility (MNRF) in the Victorian Node of the Australian National Fabrication Facility (ANFF). The \ac{SSPD} fabrication was performed at the NSW Node
and ACT Node of the NCRIS-enabled ANFF. 

All authors acknowledge the work from Ruixuan Wang and Jingwei Li, who contributed to the fabrication of the samples investigated.

\section*{Declarations}

\begin{itemize}

\item Conflict of interest/Competing interests

The authors declare that they have no competing interests.
\item Authors contribution

Conceptualization: S.C.,S.R.,J.C.McC.,Q.L.,A.L.; Data curation: J.B.,A.L.; Formal analysis: J.B.,A.L.; Investigation: J.B.,A.L.,S.Q.L.,J.McC.; Methodology: A.L., S.C., S.R.,J.McC.,Q.L.; Resources: A.L., S.C., S.R.,J.McC.,Q.L.; Supervision: S.C., B.C.J.; Visualization: J.B.,A.L.; Writing - original draft: J.B.,A.L.,S.C.; Writing - review \& editing: J.B.,A.L.,S.Q.L.,B.C.J, Q.L. All authors have read
and agreed to the published version of the manuscript.
\end{itemize}

\end{document}